\documentclass[12pt]{article}

\usepackage{latexsym}

\usepackage[font=small,labelfont=bf]{caption}

\usepackage{graphics}
\usepackage{graphicx}
\usepackage{epstopdf}
\usepackage{amssymb}
\usepackage{tabularx}
\usepackage{caption}
\usepackage{subcaption}
\usepackage{hyperref}
\usepackage{tabularx}
\usepackage{gensymb}
\usepackage{mathtools,amsmath,amssymb,mathrsfs,color,esint}
\usepackage{array}
\usepackage{makecell}
\usepackage{float}
\usepackage{dcolumn}

\begin{document}


\title {Polarized image of equatorial emission in horizonless spacetimes: naked singularities}

\author{
Valentin Deliyski$^{1}$\footnote{E-mail: \texttt{vodelijski@phys.uni-sofia.bg}}, \,
Galin Gyulchev$^{1}$\footnote{E-mail: \texttt{gyulchev@phys.uni-sofia.bg}},  \,
Petya Nedkova$^{1}$\footnote{E-mail: \texttt{pnedkova@phys.uni-sofia.bg}},
\\ Stoytcho Yazadjiev$^{1,2}$\footnote{E-mail: \texttt{yazad@phys.uni-sofia.bg}}\\ \\
 {\footnotesize${}^{1}$ Faculty of Physics, Sofia University,}\\
  {\footnotesize    5 James Bourchier Boulevard, Sofia~1164, Bulgaria } \\
    {\footnotesize${}^{2}$ Institute of Mathematics and Informatics,}\\
{\footnotesize Bulgarian Academy of Sciences, Acad. G. Bonchev 8, } \\
  {\footnotesize  Sofia 1113, Bulgaria}}
\date{}
\maketitle

\begin{abstract}
We study the linear polarization from the accretion disk around weakly and strongly naked Janis-Newman-Winicour singularities. We consider an analytical toy model of thin magnetized fluid ring orbiting in the equatorial plane and emitting synchrotron radiation. The observable polarized images are calculated and compared to the Schwarzschild black hole for physical parameters compatible with the radio source M87. For small inclination angles the direct images of the weakly naked singularities closely mimic the Schwarzschild black hole. The deviation in the polarization properties increases if we consider larger inclination angles or higher order images as for indirect images the polarization intensity grows several times in magnitude compared to black holes. Strongly naked singularities produce significant observational signatures already in the direct images. They create a second image of the fluid ring with  times larger polarization intensity and characteristic twist of the polarization direction. Due to this additional structure they can be distinguished in polarimetric experiments.
\end{abstract}

\section{Introduction}

Recent advances in gravitational physics opened several observational channels for testing the properties of the gravitational interaction in the strong field regime \cite{EHT1}. This inspired a broad range of efforts for resolving the fine structure of black holes by searching for different kinds of hair that signifies for modification of the gravitational theory \cite{Amarilla:2010}-\cite{Herdeiro:2021}. Another line of research  pursues the experimental detection of more exotic compact objects which mimic black holes in their phenomenology \cite{Nedkova:2013}-\cite{Kunz:2023}. These compact objects arise naturally in the gravitational theories but typically violate some of the fundamental laws in general relativity which constrain a physically reasonable system. Such issues can be evaded by introducing modifications of the gravitational theory or considering quantum gravity effects.

A common example are the naked singularities. They arise as end-states of gravitational collapse but their formation is restricted by the cosmic censorship conjecture since they violate causality \cite{Joshi:1993}-\cite{Dey:2020}. However, in a broader picture naked singularities can be justified as artifacts of the incompleteness of general relativity. They are expected to be resolved in a modified gravitational theory which describes more adequately  the gravitational interaction at Planck scales being replaced by regular soliton-like configurations. This motivates the investigation of naked singularity spacetimes as alternatives of black holes in astrophysical context. Together with wormholes and exotic stars, they are naturally considered in the observational tests of the presence of the event horizon or the validity of the Kerr hypothesis.

Naked singularities can mimic black holes in the imaging experiments if they possess a photon sphere. In this case, their shadows and accretion disk images can be indistinguishable from black holes with the current resolution of the Event Horizon Telescope (EHT) \cite{Shaikh:2019}-\cite{Kocherlakota:2021}. However, there exist further classes of naked singularities which demonstrate characteristic signatures. If the spacetime contains no photon sphere or possesses more complicated structure of stable and unstable light rings, a reflective barrier may form in the neighbourhood of the singularity. Thus, a region of repulsive gravitation field arises in its vicinity which  light cannot access and scatters away. This situation is reflected observationally by the absence of shadow and the formation of multi-ring structure at the center of the accretion disk images \cite{Nedkova:2020}, \cite{Nedkova:2021}.

In this work we focus on naked singularities which demonstrate similar lensing properties as the Schwarzschild black hole. They possess a single unstable light ring and the size of their shadow deviates only slightly from the Schwarzschild black hole within the experimental error of EHT.  In such cases it is important to address the question whether we could distinguish the two types of spacetimes by means of imaging experiments. The polarization of the emission from the accretion disk is one of the channels which can provide further observational signatures of the underlying spacetime. It encodes independent information for the interaction of the background geometry with the magnetic field of the accreting plasma and could lead to new characteristic effects.

Recently, the Event Horizon Telescope collaboration provided the first polarized images of the radio source at the center of the nearby galaxy M87 \cite{EHT7}, \cite{EHT8}. They contain a certain fraction of linearly polarized emission which is interpreted as produced by synchrotron radiation taking into account the properties  of the radio source. Performing general relativistic magnetohydrodynamic (GRMHD) simulations the observable polarization was calculated for a range of physical models of the accreting plasma showing that the observational data is mostly consistent with the predictions of the magnetically arrested disk (MAD) models.

In addition a simple analytical toy model was constructed which represents a thin ring of magnetized fluid orbiting in the equatorial plane of the Schwarzschild black hole \cite{Narayan:2021} further extended for the Kerr black hole in \cite{Gelles:2021}. It depends on a few physical parameters with a clear meaning and yet it manages to reproduce the observational data  with a reasonable precision. This result allows to make predictions for the observable polarization based on analytical treatment and establish more transparent correlations between the physical variables in the parametric space and the  features of the observable image.

Performing such investigations it was demonstrated that the structure of the polarized images is mainly determined by the distribution of the magnetic field \cite{Palumbo:2021}, \cite{Emami:2023}. The effect of the fluid velocity and the angular momentum of the central compact object are subdominant and the spin can be estimated only indirectly by means of its correlation with the magnetic field pattern which reproduces best the polarization data. 

Estimating the sensitivity of the polarization structure to the physical settings it is important to consider the influence of the spacetime geometry. In this study we address this issue by applying the analytical model of the synchrotron emitting equatorial ring to naked singularity spacetimes. We choose a simple class of static spacetimes representing the unique spherically symmetric solution to the Einstein-scalar field equations mostly famous as the Janis-Newman-Winicour naked singularity. Then, we simulate the observable polarization investigating its dependence on the physical parameters of the model and comparing to the Schwarzschild black hole. Our aim is to access the potential of the linear polarization as a probe of the underlying spacetime and make conclusions how effectively we can use it as an observational signature for determining the nature of the central compact object. The study is a part of a larger project on the properties of the observable polarization in horizonless spacetimes following our previous work on traversable wormholes \cite{Nedkova:2023}. The polarization for black holes in the modified theories of gravity or interacting with external fields was studied in a similar context in \cite{Qin:2021}-\cite{Zhang:2022}, while various theoretical features of the polarized emission in the Kerr spacetime were recently investigated in \cite{Lupsasca:2020}-\cite{Himwich:2020} building up on the classical works \cite{Bardeen:1972}-\cite{Agol:1997}.

The paper is organized as follows. In the next section we briefly describe the Janis-Newman-Winicour naked singularities and its relevant features. In section 3 we review the analytical model of the magnetized equatorial fluid ring  emitting synchrotron radiation. We further describe the computational procedure  which we apply for obtaining the observable polarization. In section 4 we present our simulated polarized images exploring different configurations of the parametric space. In particular we study the influence of the magnetic field orientation, the inclination angle of the observer and the order of the lensed image. We superpose the observable polarization for the naked singularities with its structure for the Schwarzschild black hole evaluating the deviation between the two spacetimes. In the last section we present our conclusions.

\section{Janis-Newman-Winicour naked singularity}

The Janis-Newman-Wincour (JNW) naked singularity represents the unique  static spherically symmetric solution to the Einstein-scalar field equations. The solution was obtained originally by Fisher \cite{Fisher:1948} and rediscovered in the later works \cite{Janis:1968}-\cite{Virbhadra:1997}. It is mostly famous in the form derived by Janis, Newman and Winicour 

\begin{equation}
ds^2 = -\left(1-\frac{2M}{\gamma r}\right)^{\gamma} dt^2
      +\left(1-\frac{2M}{\gamma r}\right)^{-\gamma} dr^2
      + \left(1-\frac{2M}{\gamma r}\right)^{1-\gamma} r^2\left(d\theta^2 + \sin^2\theta d\phi^2\right),
\end{equation}
where the scalar field takes the form
\begin{equation}
\varphi = \frac{q\gamma}{2M} \ln\left(1-\frac{2M}{\gamma r}\right).
\end{equation}
The solution is determined by its ADM mass $M$ and the scalar charge $q$. The parameter $\gamma$ takes the range $\gamma\in(0,1)$ and characterizes the charge to mass ratio 

\begin{eqnarray}\label{param}
\gamma = \frac{M}{\sqrt{M^2+q^2}}.
\end{eqnarray}
In the limit $\gamma=1$ the scalar charge vanishes and we recover the Schwarzschild black hole.

\begin{figure*}[h!]
    \centering
    \begin{subfigure}[t]{0.8\textwidth}
        \includegraphics[width=\textwidth]{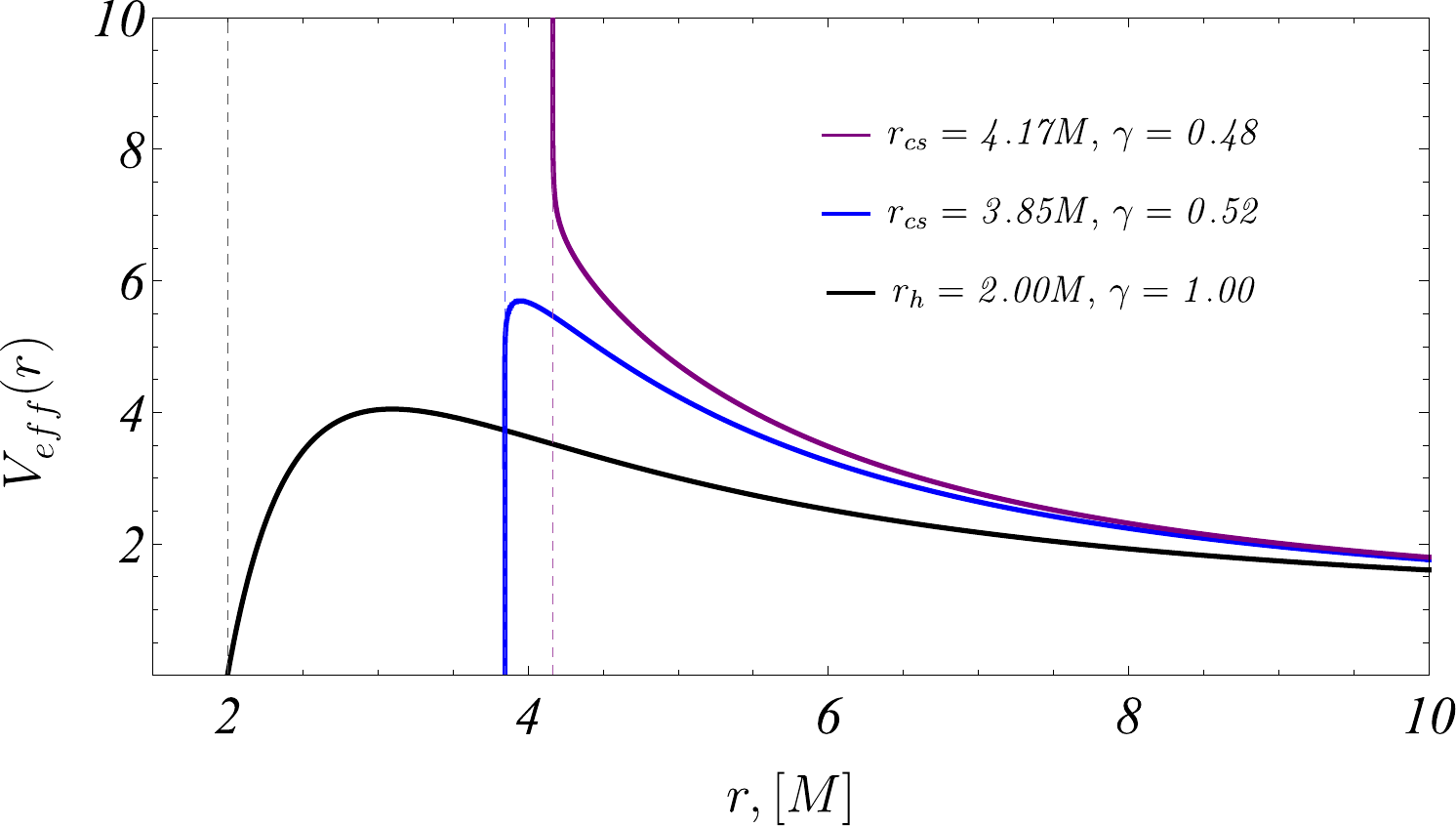}
           \end{subfigure}
           \caption{\label{fig:V_eff}\small Behaviour of the effective potential for the null geodesics for weakly naked singularities (blue solid line), strongly naked singularities (purple solid line), and the Schwarzschild black hole (black solid line). We represent by  a dashed line in the corresponding color the location of the curvature singularity $r_{cs}$ for the naked singularities, and the event horizon $r_h$ for the Schwarzschild black hole. We choose values of the scalar field parameter $\gamma = 0.52$ and $\gamma = 0.48$ for the weakly and strongly naked singularities, respectively.}
\end{figure*}

The Janis-Newman-Winicour naked singularity is characterized by two regimes with respect to its lensing properties \cite{Virbhadra:1998}-\cite{Gyulchev:2008}. For values of the scalar field parameter in the range $\gamma\in(0.5,1)$ it contains a photon sphere and its lensing properties are similar to those of the Schwarzschild black hole. This class of solutions are called weakly naked singularities. For scalar fields $\gamma\in(0,0.5)$ the solutions possess no light rings and the effective potential for the null geodesics diverges in the vicinity of the singularity \cite{Nedkova:2020} (see Fig. $\ref{fig:V_eff}$ ). This behavior leads to distinctive observational feature such as absence of shadow and appearance of multi-ring structure at the center of the accretion disk images. Solutions in this range of scalar fields the solutions are classified as strongly naked singularities.

The location of the photon sphere for the JNW naked singularity are given by the expression

\begin{equation}\label{ph_sphere}
r_{ph} = (2\gamma + 1)M/\gamma,
\end{equation}
while the marginally stable circular timelike orbits correspond to the solutions of the equation \cite{Nedkova:2020}

\begin{equation}\label{ISCO_eq}
r^2\gamma^2 - 2r\gamma(3\gamma + 1) + 2(2\gamma^2 + 3\gamma + 1) =0.
\end{equation}

For $\gamma =0.5$ the photon sphere disintegrates and Eq. ($\ref{ISCO_eq}$) no longer possesses a unique real solution. Therefore, for a certain range of scalar field $\gamma<0.5$ the stable circular particle orbits are located in two disconnected regions consisting of an inner annulus are and outer disk which extends to infinity.

In this work we concentrate mainly on the weakly naked singularities and compare their linear polarization properties with the Schwarzschild black hole. We will further restrict the scalar field range to $\gamma\in[0.53,1)$ since for these values the apparent size of the shadow for the JNW solution is compatible with the EHT observations of M87*. In the last section we will briefly investigate strongly naked singularities. They give rise to distinctive polarisation effects which may be useful in understanding future galactic targets.

\section{Polarized image of an equatorial emitting ring}

In this work we consider the physical model of a thin ring of fluid orbiting in the equatorial plane around the Janis-Newman-Winicour naked singularity. The fluid rotates at a fixed coordinate radius in a constant local magnetic field  emitting linearly polarized synchrotron radiation. The observable polarization at large distances is calculated by parallel-transporting the polarization vector along the null geodesics and obtaining its projection on the observer's sky.

The model was already described for the Schwarzschild and Kerr black holes in \cite{Narayan:2021}, \cite{Gelles:2021} and presented for a general static spherically symmetric spacetime in \cite{Nedkova:2023}. Therefore, we outline here only the basic steps referring to previous works for details. In order to obtain  the polarization in the fluid rest frame, we introduce an orthonormal tetrad $\{e^{\mu}_{(a)}\}$

\begin{eqnarray}\label{eq:tetrad}
     &&e_{(t)} = \left(1-\frac{2M}{\gamma r}\right)^{-\gamma/2}\partial_t, \,\,\,\,\,
    e_{(r)} = \left(1-\frac{2M}{\gamma r}\right)^{\gamma/2}\partial_r, \,\,\,\,\,
    e_{(\theta)} =  \frac{1}{r} \left(1-\frac{2M}{\gamma r}\right)^{(\gamma-1)/2}\partial_{\theta}, \nonumber \\[2mm]
    &&e_{(\phi)} = \frac{1}{r\sin\theta}\left(1-\frac{2M}{\gamma r}\right)^{(\gamma-1)/2}\partial_{\phi},
\end{eqnarray}
and assume that the fluid moves with velocity $\vec{\beta}$ in the $(r)$-$(\phi)$ plane of the local frame. We can further parameterize the fluid velocity as

\begin{equation} \label{eq:betaboost}
\vec{\beta} = \beta\left(\cos\chi\,(r)+\sin\chi\,(\phi)\right).
\end{equation}
by means of its magnitude $\beta$ and local direction $\chi$. The local rest frame attached to
the boosted emitter is obtained by the standard Lorenz transformation from the orthonornal frame $\{e^{\mu}_{(a)}\}$

\begin{eqnarray}\label{Lorenz_tr}
\hat{e}^{\,\mu}_{(a)} = \Lambda^{\hspace{0.3cm}(b)}_{(a)}e^{\mu}_{(b)},
\end{eqnarray}

\begin{eqnarray}
    \small
    &&\Lambda
        =\begin{pmatrix}
             \gamma  & -\beta  \gamma  \cos \chi & -\beta  \gamma  \sin \chi & 0 \\
             -\beta  \gamma  \cos \chi & (\gamma -1) \cos ^2\chi+1 & (\gamma -1) \sin \chi \cos \chi & 0 \\
             -\beta  \gamma  \sin \chi & (\gamma -1) \sin \chi \cos \chi & (\gamma -1) \sin ^2\chi+1 & 0 \\
             0 & 0 & 0 & 1 \\
        \end{pmatrix}, \\[2mm]
    \normalsize
    &&\gamma =(1-\beta^2)^{-1/2}
\end{eqnarray}

We assume that the fluid is located in a constant local magnetic field and denote its components in the fluid rest frame by $\vec{B} =\left(\hat{B}^{ r}, \hat{B}^{\phi}, \hat{B}^{\theta}\right)$. Considering the local 3-momentum  by $\vec{p} =\left(\hat{p}^{ r}, \hat{p}^{\phi}, \hat{p}^{\theta}\right)$,  we can express the polarization of the synchrotron radiation emitted by the rotating fluid as

\begin{equation} \label{eq:polcross}
  \vec{f} = \frac{\vec{p} \times \vec{B}}{|\vec{p}|}.
\end{equation}
where $f^{(t)}=0$. We further normalize the polarization vector as

\begin{equation}
\hat{f}^{a}\hat{f}_{a} = \sin^2\zeta |\vec{B}|^2,
\end{equation}
where $\zeta$ is the angle between the fluid's 3-momentum and the magnetic field

\begin{equation} \label{zeta_B}
\sin \zeta = \frac{| \vec{p} \times \vec{B}|}{|\vec{p}||\vec{B}|}.
\end{equation}

Using the Lorenz transformation ($\ref{Lorenz_tr}$), we can recover the components of the polarization 4-vector $f^{\mu}$ in the spacetime coordinates in the standard way.

In order to obtain the observable polarization, we should calculate its transformation while the electromagnetic radiation propagates from the emission point to the asymptotic observer. The polarization 4-vector is parallel transported along the null geodesics remaining normal to the 4-momentum $p^{\mu}$, i.e. it satisfies

\begin{eqnarray}
&&p^\mu \nabla_{\mu} p_\nu =0 \nonumber \\[2mm]
&&p^{\mu}\nabla_{\mu}f_\nu=0, \;\;\; p^\mu f_\mu=0.
\end{eqnarray}

In static spherically symmetric spacetime these differential equations can be solved by purely algebraic manipulations due to the hidden symmetries which lead to the presence of irreducable Killing and Killing-Yano tensors \cite{Nedkova:2023}. In particular, there exists a second order Killing-Yano tensor $Y_{\mu\nu}$, which possesses the following non-zero components for the Janis-Newman-Winicour spacetime

\begin{eqnarray}
Y_{\theta\phi}=-Y_{\phi\theta}=R^3(r) \sin\theta.
\end{eqnarray}

It further generates a conformal Killing-Yano tensor  ${\tilde Y}_{\mu\nu}$ given by the Hodge dual, i.e. $ {\tilde Y}_{\mu\nu} = \frac{1}{2}\epsilon_{\mu\nu\alpha\beta}Y^{\alpha\beta}$. These symmetries give rise to two additional constants of motion $\kappa_1$ and $\kappa_2$ along the null geodesics

\begin{eqnarray}
\kappa_1= \frac{1}{2} {\tilde Y}_{\mu\nu}p^\mu f^\nu , \;\;\;    \kappa_2= Y_{\mu\nu} p^\mu f^\nu,
\end{eqnarray}
which are connected to the the photon's 4-momentum $p^{\mu}$ and polarization vector $f^{\mu}$ as

\begin{eqnarray}\label{kappa_12}
&&\kappa_1= R(r) e^{\nu(r) +\lambda(r)} (p^t f^r - p^rf^t), \nonumber  \\[2mm]
&&\kappa_2=  R^3(r) \sin \theta (p^\theta f^\phi - p^\phi f^\theta).
\end{eqnarray}

In this way we obtain algebraic expressions for the polarization 4-vector which are valid at any point of the spacetime allowing us to determine the observable polarization solely by means of the boundary data on the photon trajectory. In order to define the observable quantities we should introduce celestial coordinates on the observer's sky. For the purpose we consider the orthonormal tetrad ($\ref{eq:tetrad}$) at the observer's position. The projection $p^{(a)}$ of the 4-momentum  in the local frame determines two angles $\alpha$ and $\beta$  \cite{Bardeen}

\begin{eqnarray}
\sin\alpha = \frac{p^{(\theta)}}{p^{(t)}}, \quad \,\,\, \tan\beta = \frac{p^{(\phi)}}{p^{(r)}},
\end{eqnarray}
which can serve as coordinates on the observer's sky. For an asymptotic observer we perform the rescaling $x =  r \alpha$ and $y=  r\beta$ and obtain the celestial coordinates

\begin{eqnarray}\label{obs_coord}
x &=& -R(r)p^{(\phi)} = -\frac{p_{\phi}}{\sin\theta_0}, \nonumber \\[2mm]
y &=&  R(r)p^{(\theta)} = p_{\theta},
\end{eqnarray}
where $\theta_0$ is the inclination angle of the observer.

Then we can express the integrals of motion $\kappa_1$ and $\kappa_2$ in terms of  the projection
of the 4-momentum and the polarization vector in the observer's frame and after some algebraic manipulations we
obtain the components of the polarization vector 

\begin{eqnarray}
f^x = \frac{x\kappa_1  + y\kappa_2}{x^2 + y^2} , \nonumber \\[2mm]
f^y = \frac{y\kappa_1 - x\kappa_2}{x^2 + y^2} ,
\end{eqnarray}
on the asymptotic observer's screen. It is further convenient to normalize the observable polarization vector to unity as

\begin{eqnarray}\label{obs_polarization}
f^x_{obs} = \frac{x\kappa_1  + y\kappa_2}{\sqrt{(\kappa_1^2 + \kappa_2^2)(x^2 + y^2)}} , \nonumber \\[2mm]
f^y_{obs} = \frac{y\kappa_1  - x\kappa_2}{\sqrt{(\kappa_1^2 + \kappa_2^2)(x^2 + y^2)}} .
\end{eqnarray}

According to the derived expressions the polarization at any point $(x,y)$ on the observer's sky is determined only by the integrals of motion $\kappa_1$ and $\kappa_2$ on the photon trajectory reaching it. On the other hand, considering eqs. ($\ref{kappa_12}$) we can calculate $\kappa_1$ and $\kappa_2$ by means of the initial data on the geodesics, i.e. by the physical quantities at the photon's emission point. Thus, we obtain 

\begin{eqnarray}\label{kappa_fr}
\kappa_1= \gamma R(r_s) \left[{\hat p^{(t)}}{\hat f^{(r)}}  + \beta \left({\hat p^{(r)}}{\hat f^{(\theta)}} - {\hat f^{(r)}}{\hat p^{(\theta)}} \right) \right], \nonumber \\[2mm]
\kappa_2 = \gamma R(r_s) \left[\left({\hat p^{(\theta)}}{\hat f^{(\phi)}} - {\hat f^{(\theta)}}{\hat p^{(\phi)}} \right)  - \beta {\hat p^{(t)}}{\hat f^{(\phi)}}\right],
\end{eqnarray}
expressing the integrals of motion in terms of the 3-momentum and the polarization vector in the local rest frame of the fluid. As a result eqs. ($\ref{obs_coord}$)-($\ref{kappa_fr}$) completely determine the observable polarization in terms of the local properties of the synchrotron emission model and the boundary data on the photon trajectory.

In our study of the polarization structure we will be interested in two observables - the polarization intensity and the direction of the polarization vector on the observer's sky. The polarization intensity is
proportional to the norm of the observable polarization vector modified by phenomenological factors which follow from the properties of the synchrotron radiation \cite{Narayan:2021}. The intensity of the synchrotron emission at the fluid's rest frame is proportional to the angle $\zeta$  between the direction of the magnetic field and the 4-velocity (see eq. ($\ref{zeta_B}$)), the frequency of the emitted photons  and the geodesic path length $l_p$ in the emitting region

\begin{eqnarray}
 l_p=\frac{\hat p^{(t)}}{\hat p^{(\theta)}}\,H,
\end{eqnarray}
where $H$ is the disk height and $\hat p^{(\mu)}$ is the 4-momentum in the fluid rest frame. In addition, we have to take into account the Doppler-boost while the radiation propagates in the spacetime towards the observer's location. It modulates the intensity by the factor $\delta^{3+\alpha_\nu}$, where $\delta$ is the gravitational redshift

\begin{equation}
\delta = \frac{E_{obs}}{E_s} = \frac{1}{\hat p^{(t)}},
\end{equation}
determined by the ratio of the photon's energy measured at the observer's frame $E_{obs}$ and the emitter's frame $E_s$. Explicitly, the polarization intensity is given by the expression \cite{Narayan:2021}

\begin{eqnarray}
|P| = \delta^{3+\alpha_\nu}\, l_p \, (\sin\zeta)^{1+\alpha_\nu},
\end{eqnarray}
where the spectral index $\alpha_\nu$ depends on the emission frequency and we take into account that the polarization vector at the observer's frame is normalized to unity. In our study we consider the value $\alpha_\nu=1$, which is consistent with the observations of M87 \cite{Narayan:2021}, \cite{Gelles:2021}. Then, we can define the components of the observed electric field as

\begin{eqnarray}\label{pol_EVPA}
E^x_{obs} &=& \delta^2\, l_p^{1/2}\, \sin\zeta\, f^x_{obs}, \nonumber \\[2mm]
E^y_{obs} &=& \delta^2\, l_p^{1/2}\, \sin\zeta\, f^y_{obs}, \nonumber  \\[3mm]
|P| &=& (E^x_{obs})^2 + (E^y_{obs})^2,
\end{eqnarray}
where $f^x_{obs}$ and $f^y_{obs}$ are given by eq. ($\ref{obs_polarization}$).

The direction of the polarization vector in the observer's frame is estimated by computing the electric vector position angle (EVPA) defined as

\begin{eqnarray}
EVPA = \arctan\left({-\frac{f^x_{obs}}{f^y_{obs}}}\right).
\end{eqnarray}

In our simulations the EVPA is measured with respect to the positive semi-axis $x>0$ on the observer's sky in counter-clockwise direction.

\section{Linear polarization for Janis-Newman-Winicour naked singularities}

In this section we simulate the observable polarization in the JNW naked singularity spacetime using the analytical model of the magnetized fluid ring emitting synchrotron radiation which we previously described. In practice, we integrate numerically the null geodesic equations for trajectories originating at the location of the fluid ring and obtain its lensed image on the observer's sky. The observable polarization at each point of the image is calculated according to eq. ($\ref{pol_EVPA}$), as the polarization intensity is proportional to the length of the polarization vector while the variation in the EVPA is encoded in its twist around the ring.

The observable polarization depends on the physical parameters of the accreting flow such as the fluid velocity and the magnetic field in the fluid's rest frame. In addition, the image is influenced by the spacetime parameters which determine the gravitational lensing. These include the parameters of the spacetime metric, the  location of the fluid ring, the inclination angle of the observer and the order of the lensed image $k$. The images are classified into direct and indirect according to the variation of the azimuthal angle $\phi$ on the photon trajectories which lead to their formation. Direct images are created by trajectories which perform no loops around the compact object before reaching the observer, i.e. they are characterized by azimuthal angles in the range $\phi\in[0,\pi)$. On the other hand, trajectories preforming $k$ half-loops around the compact object give rise to indirect images of order $k$ with variation of the azimuthal angle $\phi\in[0, (k+1)\pi]$. In this terminology direct images can be also classified as images of order $k=0$. Larger inclination angles and higher image orders are associated with stronger lensing effects and more pronounced impact of the underlying spacetime.

In this study we explore the hypothesis that naked singularities may produce polarized images which are compatible with the observational data for M87*. Therefore, we choose values of the physical parameters which reproduce the main qualitative signatures of the radio source. The observed image of M87* is characterized by a twisting polarization pattern where the polarization vector rotates up to $70^\circ$ around the ring. In addition , the most intensive flux is concentrated on the right-hand side of the image and gradually declines towards the north and the south poles. Previous studies exploring the magnetized ring model in the Schwarzschild background  provided intuition about the impact of the various parameters \cite{Narayan:2021}. The polarization pattern is determined primarily by the direction of the magnetic field leading to fairly strong constraints on its direction. Thus, in order to reproduce the observed polarization of M87* in the Schwarzschild spacetime, vertical magnetic fields should be ruled out since they causes contradictory distribution of the polarization intensity. The magnetic field should be  restricted entirely to the equatorial plane possessing both radial and azimuthal components in order to generate the proper twist of the polarization vector around the image. For the Schwarzschild black hole it was estimated that the best fit of the observational data is achieved for magnetic field direction $B= [B_r, B_\phi, B_\theta] = [0.87, 0.5, 0]$. 

Considering purely equatorial magnetic fields reduces the number of independent parameters of the model. In such configurations it is physically most reasonable that the magnetic field follows the flow of the fluid velocity. Consequently, we can assume that the angles $\eta$ and $\xi$ which determine their directions are related as $\eta =\xi$ or $\eta = \pi + \xi$ as the choice of parallel or anti-parallel fields does not influence the polarization pattern.

In the next section we will probe how these conclusions are influenced by modifying the spacetime geometry by considering naked singularities. We choose small inclination angle close to the estimated inclination angle $\theta = 17^\circ$ for the radio source M87*. The location of the fluid ring is selected in the region of strong gravitational interaction corresponding to the ISCO for the Schwarzschild black hole $r=6M$ or  to $r=4.5M$ which provide the apparent size of the emission ring according to the EHT observations of M87*. Small variations of the radius of the fluid ring within the strong gravity region are not essential since they does not influence considerably the polarization pattern \cite{Narayan:2021}, \cite{Nedkova:2023}. 

 \subsection {Direct images for weakly naked singularities}

We begin our analysis by testing qualitatively the structure of the polarization pattern in naked singularity spacetimes. In particular, we need to confirm whether the restrictions for purely equatorial magnetic field are still valid in the new geometry. The characteristic pattern in vertical magnetic field, which is incompatible with the observations for the Schwarzschild spacetime results from the interplay of abberation and lensing effects \cite{Narayan:2021}. While the impact of the abberation  will appear consistently for any static spacetime, the strong gravitational lesning for another compact object may counteract and lead to a different symmetry of the polarized image.

For the purpose we simulate the observable polarization for the Janis-Newman-Winicour naked singularity in purely vertical magnetic field. We consider two emission radii of the fluid ring at $r=6M$ and $r=4.5M$ and compare the resulting images with the Schwarzschild black hole for the range of scalar field parameter $\gamma = [0.53, 1)$. As we discussed in section 2, the Janis-Newman-Winicour family of solutions converges to the Schwarzschild black hole for $\gamma=1$ while the lower limit is chosen as a result of the analysis of the compatibility of the observable shadow size for M87* with various alternatives of the Kerr black hole. Following \cite{Kocherlakota:2021} the Janis-Newman-Winicour naked singularity reproduces the EHT observational data in this respect for scalar fields in the range $\gamma = [0.53, 1)$.

\begin{figure}[t]
\centering
   \includegraphics[width=0.6\textwidth]{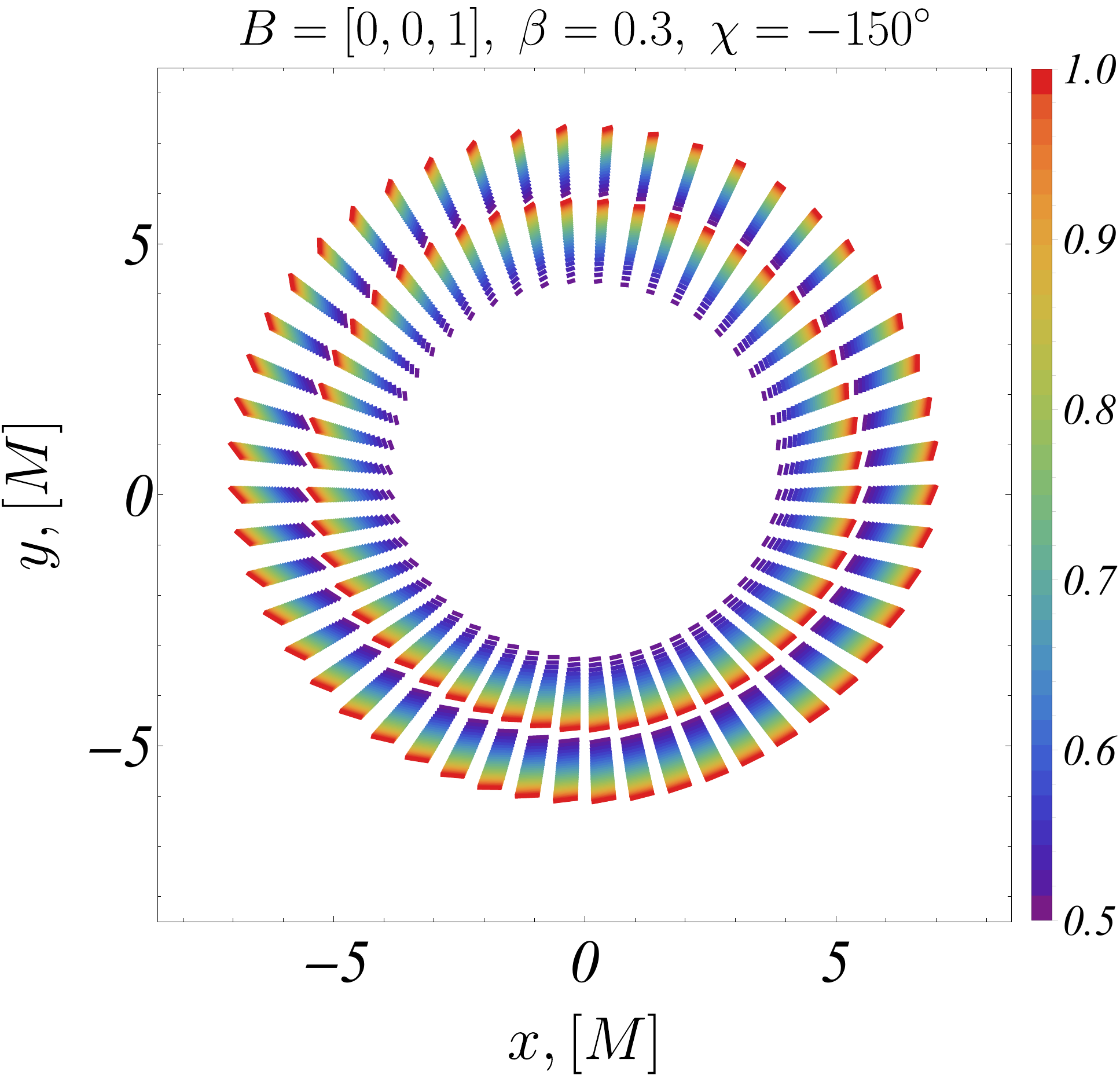}
 \caption{Polarization in vertical magnetic fields for weakly naked singularities with different scalar field parameter $\gamma$. Each color represents the observable polarization of the orbits located at $r=6M$ (outer ring) and $r=4.5M$ (inner ring) for a particular naked singularity with  $\gamma\in (0.5,1)$. The polarization for the Schwarzschild black hole corresponds to the upper limit $\gamma=1$ represented in red. The inclination angle is $\theta = 20^\circ$.}
\label{fig:pol_20v}
\end{figure}

In Fig. $\ref{fig:pol_20v}$ we present our results showing that the naked singularity spacetime produces qualitatively similar polarized images in vertical magnetic field as the Schwarzschild black hole. The polarization intensity decreases for lower values of the scalar field parameter, i.e. for greater deviations from the Schwarzschild solution. However, its distribution around the image preserves the same symmetry, the most pronounced flux being concentrated at the bottom of the ring and declining upwards. Therefore, vertical magnetic fields in the Janis-Newman-Winicour spacetime still lead to polarized images which contradict the observations of M87*.

\begin{figure}[]
\centering
   \includegraphics[width=\textwidth]{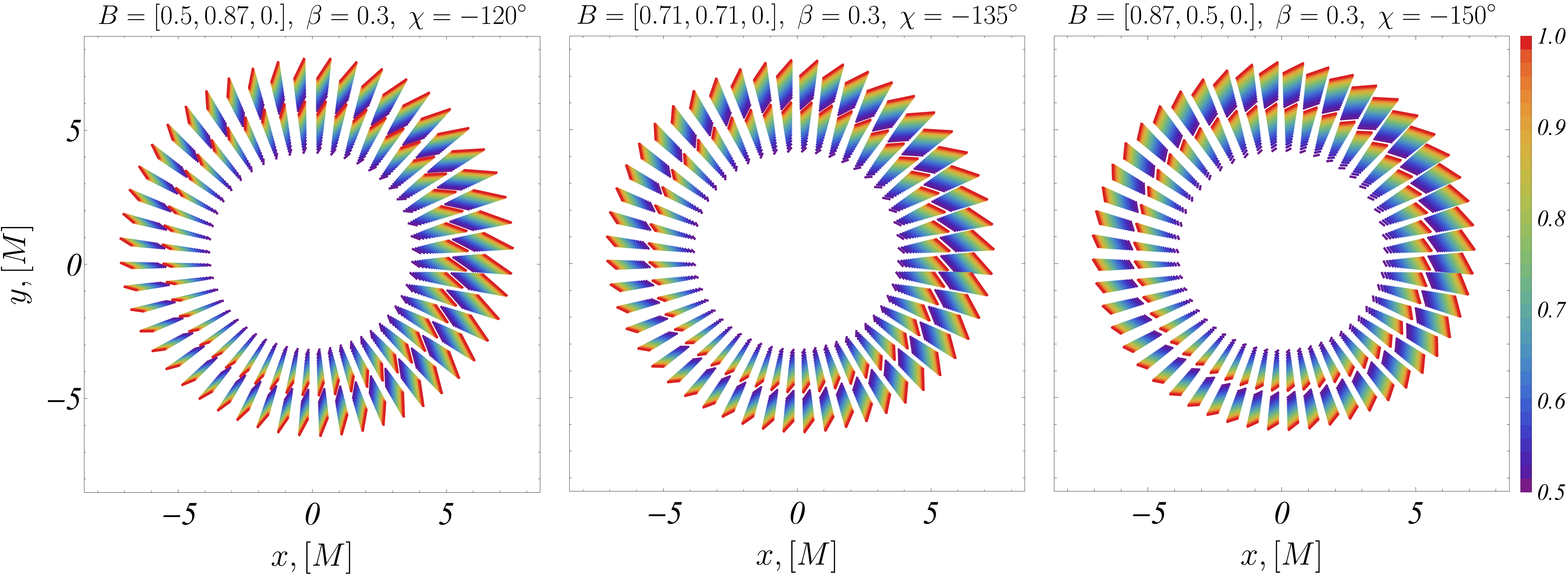}
 \caption{Polarization in equatorial magnetic field for weakly naked singularities with different scalar field parameter $\gamma$. Each color represents the observable polarization of the orbits located at $r=6M$ (outer ring) and $r=4.5M$ (inner ring) for a particular naked singularity with  $\gamma\in (0.5,1)$. The polarization for the Schwarzschild black hole corresponds to the upper limit $\gamma=1$ represented in red. The inclination angle is $\theta = 20^\circ$.}
\label{fig:pol_20}
\end{figure}

In the following analysis we impose the restriction for purely equatorial magnetic field and perform similar simulations testing the qualitative behavior of the observable polarization in this scenario. In Fig. $\ref{fig:pol_20}$ we present the polarized images for several configurations of equatorial magnetic fields showing that the twist of the polarization  vector for the naked singularity follows closely the pattern for the Schwarzschild black hole. The intensity of the polarization flux is generally lower, and decreases with the deviation from the Schwarzschild black hole. However, it retains the same distribution around the ring with peak values at the right-hand side of the image and declining towards the north and south poles.

As a result we can conclude that for small inclination angles the Janis-Newman-Winicour naked singularity reproduces the morphology of the direct polarized images for the Schwarzschild black hole. The polarization pattern is compatible with the observations of M87* for similar distribution of the magnetic field. In order to access the possibility to distinguish the two types of compact objects by polarization experiments we should perform quantitative analysis comparing the deviation of the polarization intensity and direction. For the purpose, we need to compare the observables for the same apparent radius rather than for the same emitting one. Although the polarized flux which we consider originates from different emission radii, the deviation is not substantial since the two spacetimes possess similar focusing effect for the direct images \cite{Nedkova:2019}.

\begin{figure}[t]
\centering
  \includegraphics[width=\textwidth]{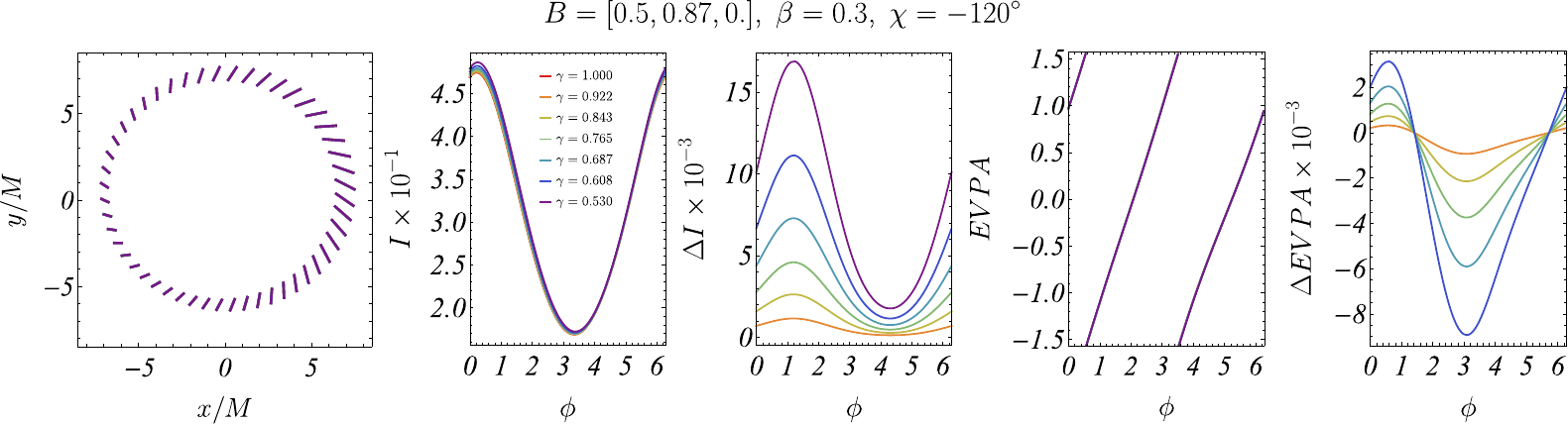}\\[2mm]
  \includegraphics[width=\textwidth]{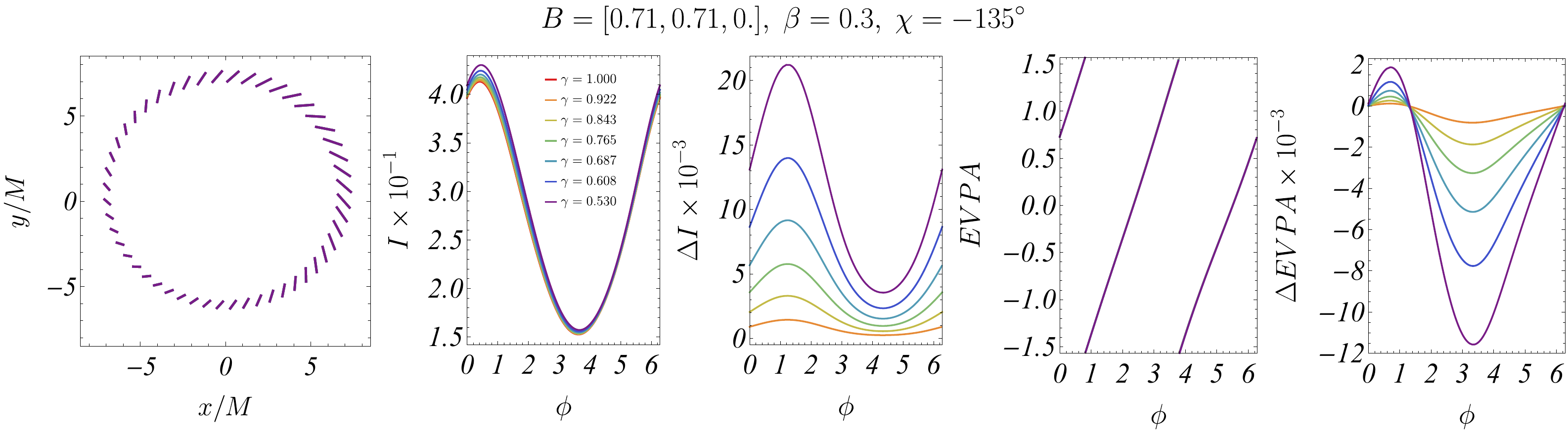}\\[2mm]
  \includegraphics[width=\textwidth]{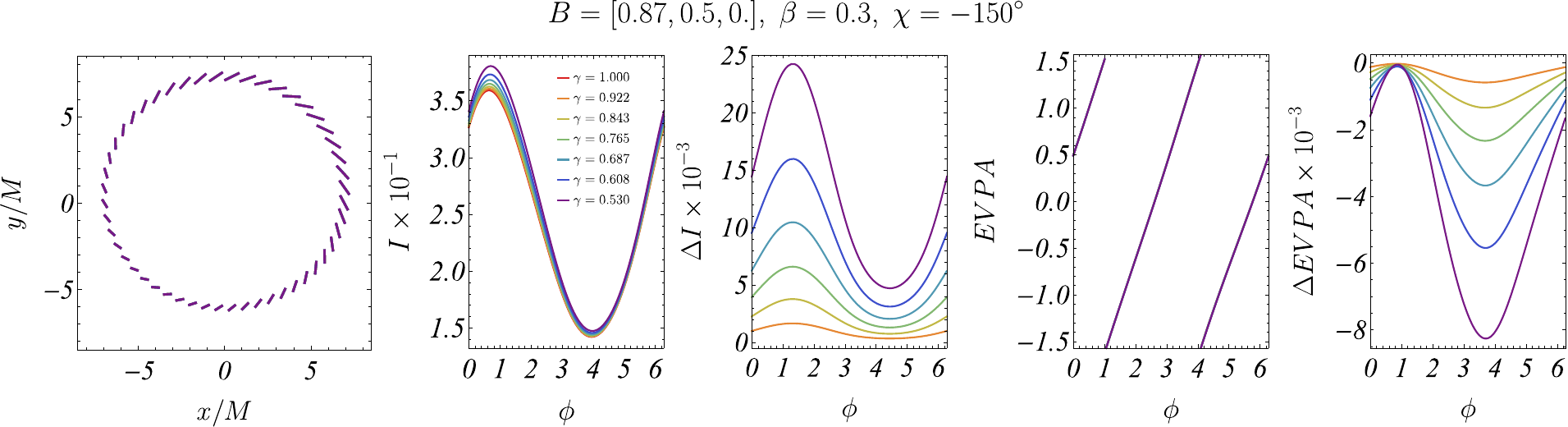}
 \caption{Linear polarization for weakly naked singularities at the inclination angle $\theta =20^\circ$. We analyze the polarization intensity I and direction EVPA as a function of the scalar field parameter $\gamma$, as well as their deviation from the Schwarzschild black hole $\Delta$I and  $\Delta$EVPA (see main text). }
\label{fig:polarization_20}
\end{figure}

Our analysis is presented in Fig. $\ref{fig:polarization_20}$ considering the polarization at the observable location of the ISCO for the Schwarzschild black hole for several naked singularity spacetimes. For each value of the scalar field parameter $\gamma$ we plot the variation of the polarization intensity and its direction around the ring as a function of the azimuthal coordinate. We further evaluate the deviation from the Schwarzschild black hole by computing the relative intensity $\Delta \text{I} = \text{I}_\text{NS} - \text{I}_\text{Sch}$ and the relative direction angle $\Delta \text{EVPA} =\text{ EVPA}_\text{NS} - \text{EVPA}_\text{Sch}$ at each point of the image. The  analysis confirms that for each $\gamma$ the intensity and the direction angle follow the same profile around the ring  as for the Schwarzschild solution with their peaks and minima located at similar values of the azimuthal angle. Since the Schwarzschild black hole represents an upper limit for the Janis-Newman-Winicour class of solutions, the polarization properties for the naked singularities can be infinitesimally close to those of black holes. The discrepancy grows continuously when $\gamma$ decreases reaching its maximum for the lower limit $\gamma =0.53$. In Table $\ref{table:theta20}$  we present the values of the maximum deviation in the polarization intensity and direction which is reached in the image for $\gamma =0.53$ in the various equatorial magnetic field configurations, which we consider. They represent an estimate for the maximum deviation in the polarization properties possible for a JNW naked singularity with shadow size compatible with the observations of M87*.

We further observe that the maximum deviation in the polarization structure for naked singularities and black holes depends on the magnetic field configuration. The discrepancy in the polarization intensity increases when the radial component of the magnetic field grows, while the twist of the polarization vector becomes more similar. However, for the values of the scalar charge which we consider none of the deviations is significant enough to be detected with the current resolution of EHT. The deviation in the polarization intensity is less than $7.5\%$ for any of the magnetic field configurations which we consider,  while the polarization vector direction differs less than $1.5\%$. For the magnetic field direction $B= [0.87, 0.5, 0]$ which provides the  best fit for the observed polarized images of M87* when the spacetime is modelled by the Schwarzschild black hole, the naked singularities produce practically identical twist of the polarization vector around the ring (max$\Delta \text{EVPA}/\text{EVPA}_\text{Sch} < 0.7\%$) for any of the considered values of $\gamma$.

These observations demonstrate that conclusions about the magnetic field configuration in the accretion flow in M87*, which are based on polarization measurements may depend weakly on the spacetime geometry at least for a certain class of ultracompact objects with compatible apparent shadow size.

\begin{table}
    \centering
      \begin{tabular}{||c|c|c|c||}
       \hline
         \thead{ Magnetic field }   &  \thead{$\left(\frac{\text{max}\,\Delta \text{I}}{\text{I}_\text{Sch}} \, [\%], \, \phi \, [rad]\right)$} & \thead{$\left(\frac{\text{max}\,\Delta \text{EVPA}}{\text{EVPA}_\text{Sch}} \, [\%] , \, \phi \, [rad]\right)$} & $\gamma$
          \\  \hline

           \thead{\vspace{0.1mm}$\text{B = [0.5, 0.87, 0]}$\vspace{0.1mm}  }   &  \thead{(4.385, 0.385$\pi$)} & \thead{(1.221, 0.986$\pi$)} & \thead{0.530}
          \\  \hline

          \thead{\vspace{0.1mm}$\text{B = [0.71, 0.71, 0]}$\vspace{0.1mm}} &  \thead{(5.928, 0.397$\pi$)} & \thead{(1.076, 1.066$\pi$)} & \thead{0.530}
          \\  \hline

          \thead{\vspace{0.1mm}$\text{B = [0.87, 0.5, 0]}$\vspace{0.1mm}} &  \thead{(7.331, 0.421$\pi$)} & \thead{(0.716, 1.178$\pi$)} & \thead{0.530}
          \\  \hline          
       \end{tabular}
      \caption{Deviation of the polarization intensity and direction for the weakly naked singularities from the Schwarzschild black hole. In each case we give the maximum relative deviations $\text{max}\,\Delta\text{I}/\text{I}_\text{Sch}$ and $\text{max}\,\Delta\text{EVPA}/\text{EVPA}_\text{Sch}$  with respect to the Schwarzschild solution, which are reached in the polarized images, and the corresponding azimuthal angle. The inclination angle is $\theta=20^{\circ}$.}
    \label{table:theta20}
\end{table}

For completeness we investigate the influence of the strong gravitational lensing on the polarization properties of the direct images. For the purpose we construct polarized images at large inclination angles following the same conventions as in Fig. $\ref{fig:polarization_20}$  These images are not supposed to be applied in modelling M87* but they are interesting from a theoretical perspective or in view of future galactic targets for the next generation EHT. 

\begin{figure}[t]
\centering
   \includegraphics[width=\textwidth]{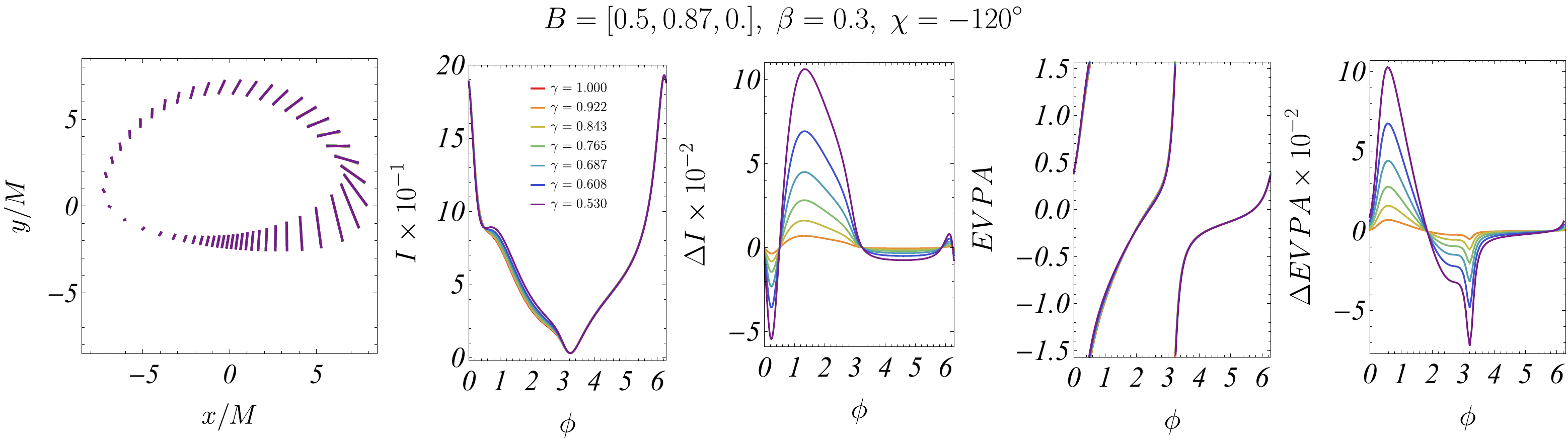}
 \\[2mm]
  \includegraphics[width=\textwidth]{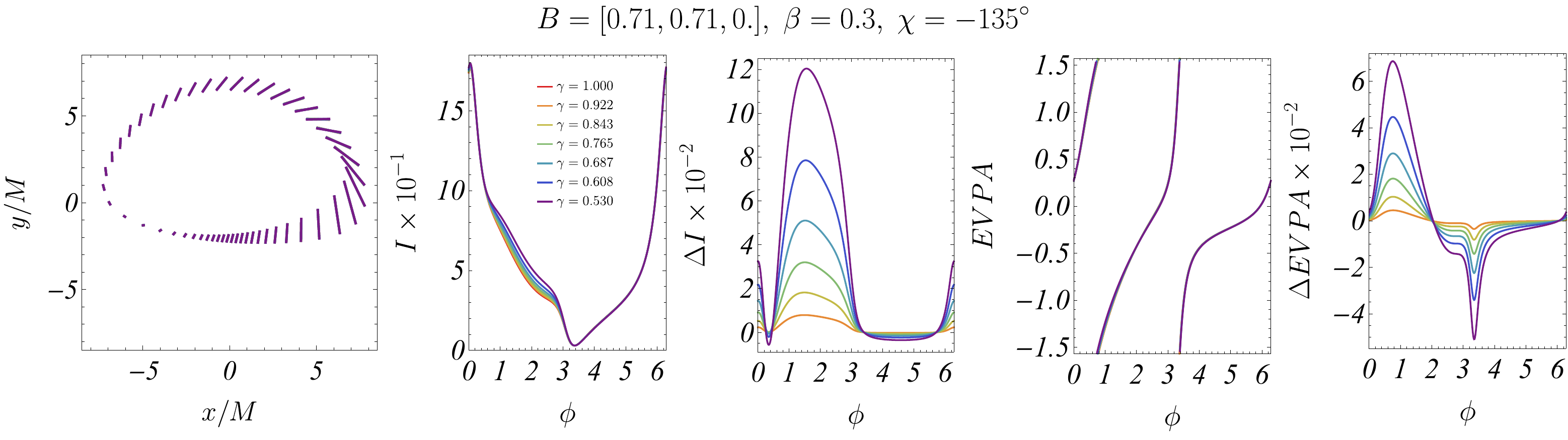} \\[2mm]
  \includegraphics[width=\textwidth]{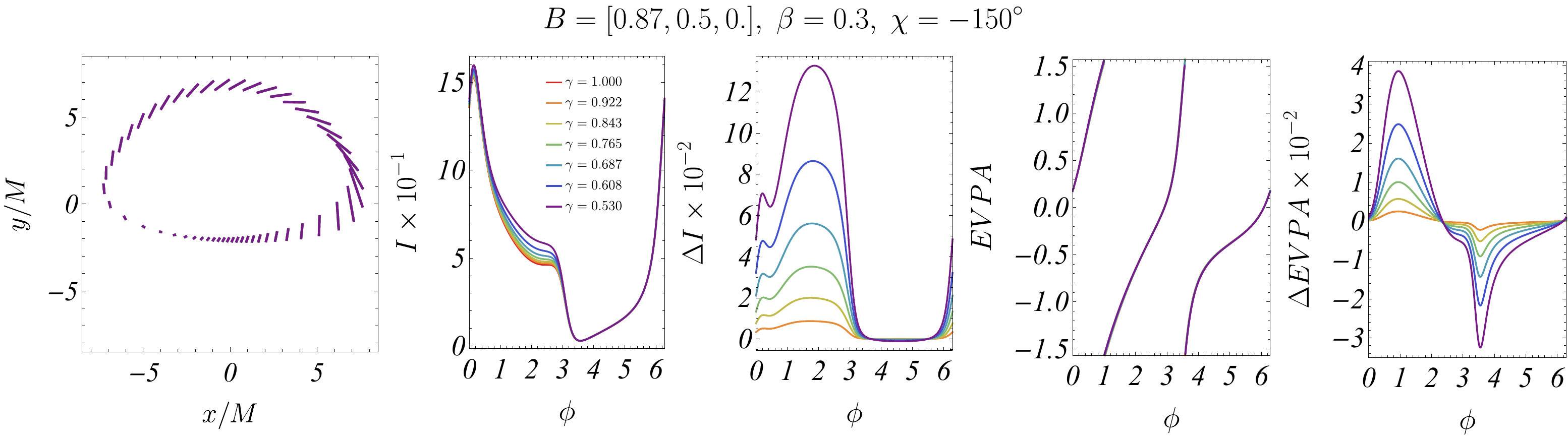}
 \caption{Linear polarization for weakly naked singularities at the inclination angle $\theta =70^\circ$. We analyze the polarization intensity I and direction EVPA as a function of the scalar field parameter $\gamma$, as well as their deviation from the Schwarzschild black hole $\Delta$I and  $\Delta$EVPA (see main text). }
\label{fig:polarization_70}
\end{figure}

In Fig. $\ref{fig:polarization_70}$  we present our results for the observable polarization at the apparent location of the ISCO for the Schwarzschild solution for inclination angle $\theta = 70\%$. The qualitative structure of the polarization pattern remains similar to the Schwarzschild black hole, however the degree of deviation in the polarization properties increases with respect to small inclination angles. In Table $\ref{table:theta70}$ we estimate the maximum deviation from the Schwarzschild black hole which is reached in the images for the scalar fields which we consider in different magnetic field configurations. The maximum deviation in the polarization intensity increases when the radial component of the magnetic field grows reaching $26.5\%$ for $B= [0.87, 0.5, 0]$. For the twist of the polarization vector we observe the opposite correlation with the magnetic field direction. The largest values of the maximum deviation in EVPA are realised for $B= [0.5, 0.87, 0]$, where we have $6.8\%$ difference in the polarization direction. Supplementary data for further values of the inclination angle is presented in Appendix A supporting these conclusions.

\begin{table}[]
    \centering
      \begin{tabular}{||c|c|c|c||}
       \hline
         \thead{ Magnetic field }   &\thead{$\left(\frac{\text{max}\,\Delta \text{I}}{\text{I}_\text{Sch}} \, [\%], \, \phi \, [rad]\right)$} & \thead{$\left(\frac{\text{max}\,\Delta \text{EVPA}}{\text{EVPA}_\text{Sch}} \, [\%] , \, \phi \, [rad]\right)$} & $\gamma$
          \\  \hline

           \thead{\vspace{0.1mm}$\text{B = [0.5, 0.87, 0]}$\vspace{0.1mm}}  &  \thead{(18.00, 0.429$\pi$)} & \thead{(6.809, 0.184$\pi$)} & \thead{0.530}
          \\  \hline

          \thead{\vspace{0.1mm}$\text{B = [0.71, 0.71, 0]}$\vspace{0.1mm}} &  \thead{(22.65, 0.497$\pi$)} & \thead{(4.482, 0.240$\pi$)} & \thead{0.530}
          \\  \hline

          \thead{\vspace{0.1mm}$\text{B = [0.87, 0.5, 0]}$\vspace{0.1mm}}  &  \thead{(26.45, 0.593$\pi$)} & \thead{(2.629, 0.301$\pi$)} & \thead{0.530}
          \\  \hline
       \end{tabular}
      \caption{Deviation of the polarization intensity and direction for the weakly naked singularities from the Schwarzschild black hole. In each case we give the maximum relative deviations $\text{max}\,\Delta\text{I}/\text{I}_\text{Sch}$ and $\text{max}\,\Delta\text{EVPA}/\text{EVPA}_\text{Sch}$  with respect to the Schwarzschild solution, which are reached in the polarized images, and the corresponding azimuthal angle. The inclination angle is $\theta=70^{\circ}$.}
    \label{table:theta70}
\end{table}

\subsection {Indirect images}

\begin{figure}[]
\centering
   \includegraphics[width=\textwidth]{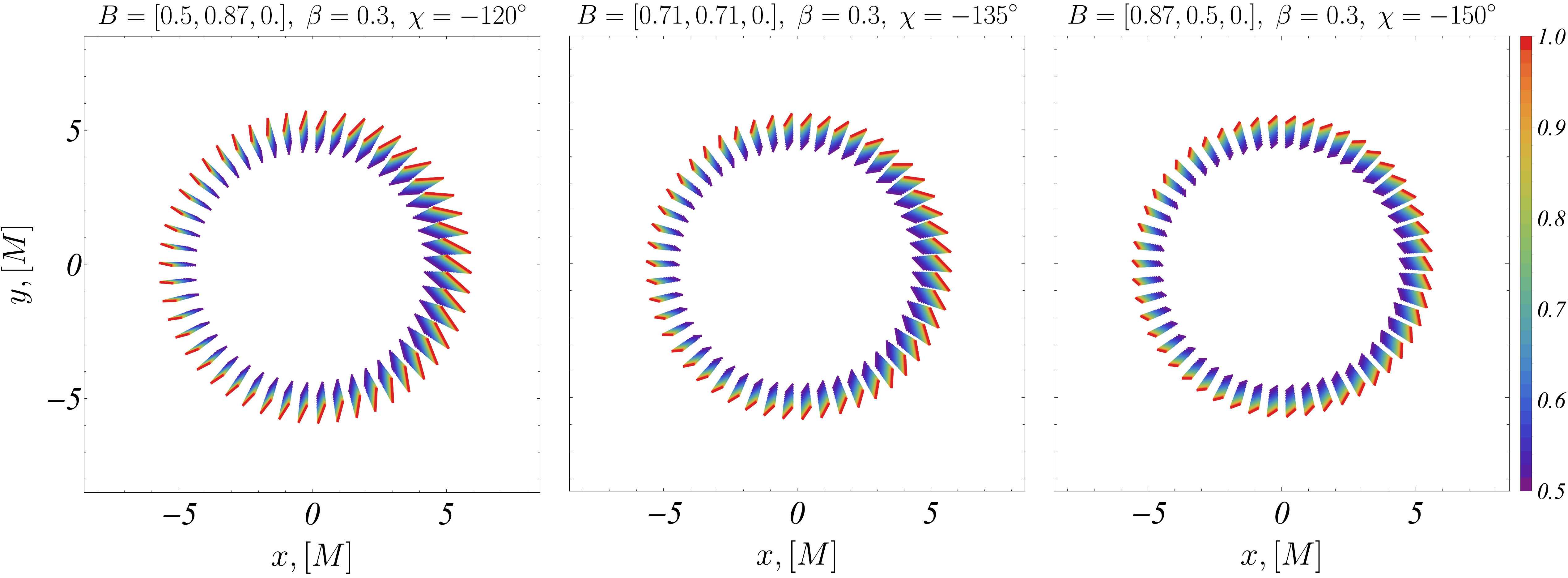}
 \caption{Indirect polarized images in equatorial magnetic field for weakly naked singularities with different scalar field parameter $\gamma$. Each color represents the observable polarization emitted at $r=6M$ for a particular naked singularity with  $\gamma\in (0.5,1)$. The inclination angle is $\theta = 20^\circ$.}
\label{fig:pol_eq_1}
\end{figure}

\begin{figure}[t]
   \centering
   \includegraphics[width=\textwidth]{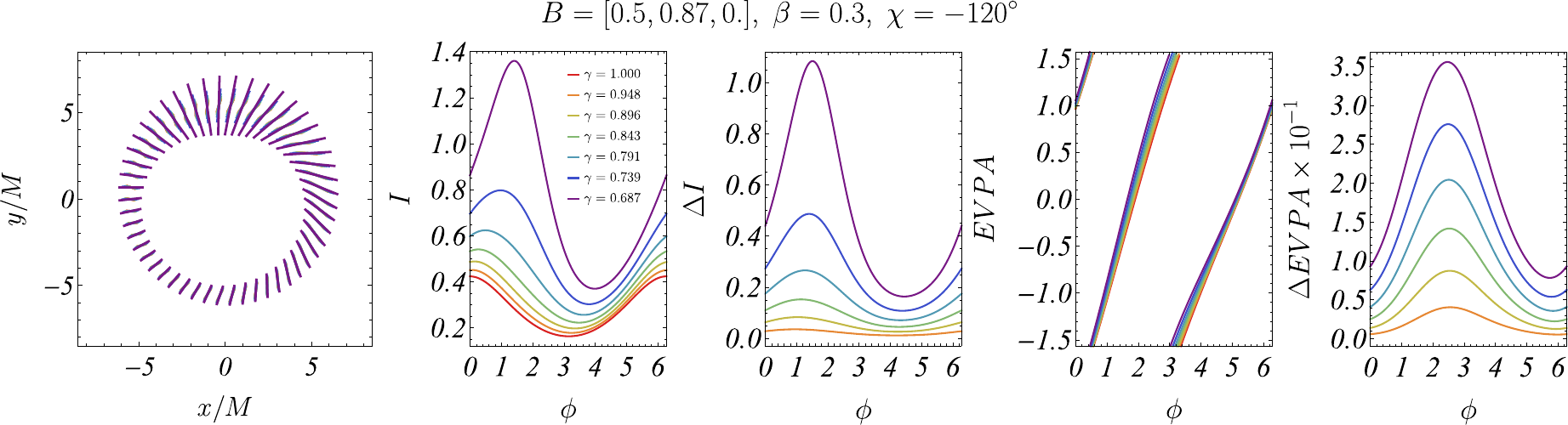}
 \\[2mm]
  \includegraphics[width=\textwidth]{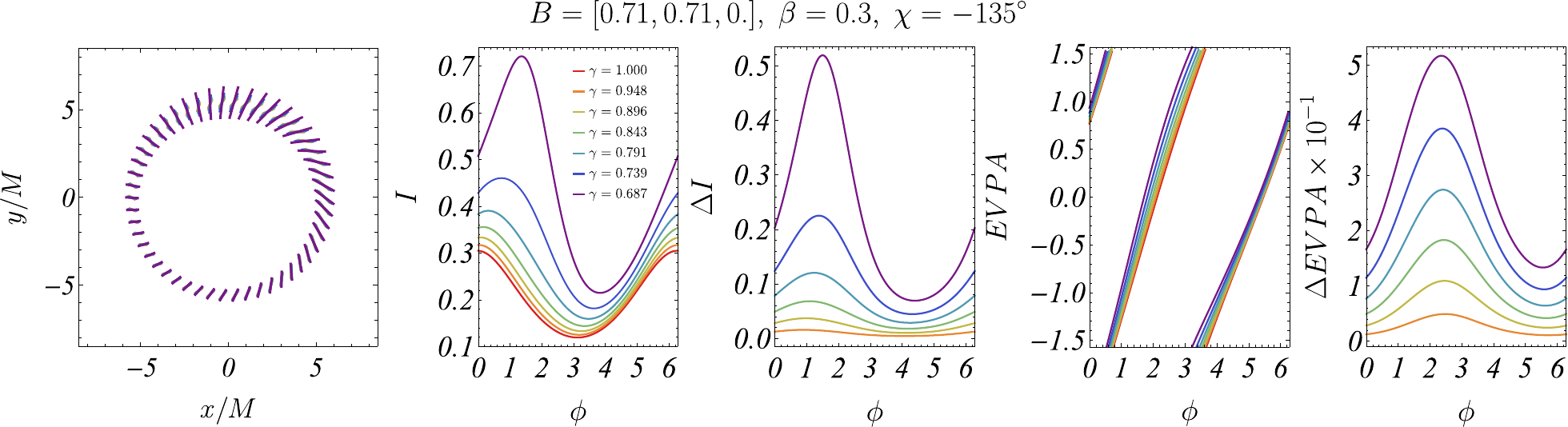} \\[2mm]
  \includegraphics[width=\textwidth]{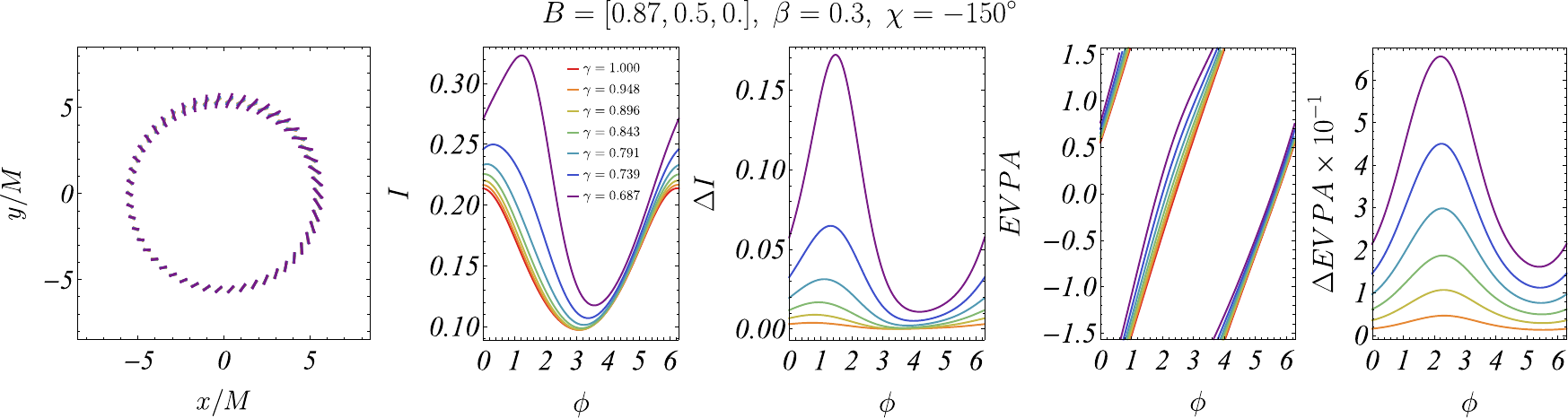}
  \caption{ Polarization of the indirect images of order $k=1$ for wormholes at the inclination angle $\theta = 20^\circ$. We analyze the polarization intensity I and direction EVPA as a function of the redshift parameter $\alpha$, as well as their deviation from the Schwarzschild black hole $\Delta$I and $\Delta$EVPA (see main text).}
\label{fig:pol1_k1}
\end{figure}

In this section we investigate the polarization of the higher order images which result from null geodesics  performing a number of half-loops around the naked singularity before reaching the observer. These images encode strong lensing effects and probe more effectively the properties of the gravitational field of the central compact object. In particular, we focus on the images of order $k=1$. In order to investigate the behavior of the indirect polarized images for different value of the scalar field parameter we calculate the observable polarization emitted at $r=6M$ for $\gamma\in(0.5,1)$ (see Fig. \ref{fig:pol_eq_1}). In order to compare the polarization properties with the Schwarzschild black hole we simulate the observable polarization at the apparent location of the ISCO for the Schwarzschild black hole produced by trajectories of order $k=1$ in the two types of spacetimes (see Fig. \ref{fig:pol1_k1}).

The strongly lensed higher order trajectories exist for a narrow range of impact parameters leading to significantly more compact images on the observer's sky than the direct trajectories. As a result, the indirect images of the thin disk around the Schwarzschild black hole and the naked singularities may possess no overlap for certain values of the scalar field parameter $\gamma$. Due to this phenomenon, for some scalar fields the impact parameters which form the indirect image of the ISCO for the Schwarzschild black hole may not correspond to any null geodesic, originating in the equatorial plane in the JNW spacetime.  Our simulations show that naked singularities with scalar field parameter in the range $\gamma\in(0.68,1)$ produce images of order $k=1$ at any point of the observable location of the ISCO for the Schwarzschild black hole. For lower values of $\gamma$ certain parts of the image possess no correspondence for naked singularities (see Appendix B).

\begin{table}
    \centering
      \begin{tabular}{||c|c|c|c||}
       \hline
         \thead{ Magnetic field }   &  \thead{$\left(\frac{\text{max}\,\Delta \text{I}}{\text{I}_\text{Sch}} , \, \phi \, [rad]\right)$} & \thead{$\left(\frac{\text{max}\,\Delta \text{EVPA}}{\text{EVPA}_\text{Sch}} , \, \phi \, [rad]\right)$} & $\gamma$
          \\  \hline

           \thead{\vspace{0.1mm}$\text{B = [0.5, 0.87, 0]}$\vspace{0.1mm}}  & \thead{(4.01, 0.481$\pi$)} & \thead{(0.55, 0.786$\pi$)} & \thead{0.687}
          \\  \hline

          \thead{\vspace{0.1mm}$\text{B = [0.71, 0.71, 0]}$\vspace{0.1mm}} &  \thead{(2.67, 0.481$\pi$)} & \thead{(1.96, 0.754$\pi$)} & \thead{0.687}
          \\  \hline

          \thead{\vspace{0.1mm}$\text{B = [0.87, 0.5, 0]}$\vspace{0.1mm}}  &  \thead{(1.19, 0.473$\pi$)} & \thead{(3.01, 0.705$\pi$)} & \thead{0.687}
          \\  \hline
       \end{tabular}
      \caption{Deviation of the naked singularities polarization from the Schwarzschild black hole for the indirect images of order $k=1$ and inclination angle $\theta=20^{\circ}$. In each case we give the maximum relative deviations $\text{max}\,\Delta\text{I}/\text{I}_\text{Sch}$ and $\text{max}\,\Delta\text{EVPA}/\text{EVPA}_\text{Sch}$  with respect to the Schwarzschild solution, which are reached in the polarized image.}
    \label{table:theta20_1}
\end{table}
Our results are presented in Fig. $\ref{fig:pol1_k1}$ where we compare the observable polarization properties for naked singularities and black holes at small inclination angles. We see that the deviation in the polarization intensity between the two types of compact objects increases substantially with respect to direct images. For the magnetic field direction $B = [0.5, 0.87, 0]$ naked singularities lead to four times more intense polarized flux than the Schwarzschild black hole (see Table $\ref{table:theta20_1}$). The deviation decreases for magnetic field configurations with dominating radial component but we still observe doube intensity for $B = [0.87, 0.5, 0]$. In addition, the intensity of the polarized flux produces a qualitatively different pattern around the ring. The location of its peak values moves upwards when the scalar field parameter $\gamma$ decreases and for $\gamma = 0.67$ the maximum intensity is concentrated at the top of the image. These signatures imply that the higher order polarized images may serve for distinguishing black holes from naked singularities if resolved in the next generation EHT.

The deviation in the polarization direction from the Schwarzschild black hole also increases substantially compared to direct images. For the magnetic field configuration $B = [0.87, 0.5, 0]$ the maximum EVPA for naked singularities is approximately 3 times larger than for black holes. It decreases when the radial component of the magnetic field decreases reaching $50\%$ for $B = [0.5, 0.87, 0]$. The distribution of the polarization direction around the ring remains qualitatively similar to the Schwarzschild solution.

\subsection{Strongly naked singularities}

In this section we consider polarization effects in the spacetime of strongly naked singularities. When the scalar field parameter reaches the range $\gamma<0.5$, the JNW singularity changes drastically its lensing properties. The spacetime contains no photon sphere as well as no stable  light rings. The effective potential diverges at the vicinity of the singularity forming a reflective barrier for photon trajectories. As a result, there exist two families of turning points for infalling null geodesics which correspond to two  disconnected ranges of impact parameters. Due to this property each equatorial circular orbit possesses a double direct image on the observer's sky \cite{Nedkova:2020}. It further leads to the formation of double images of the accretion disk as the second image represents a nested ring-like structure at small celestial angles.

\begin{figure}[t]
\centering
  \includegraphics[width=\textwidth]{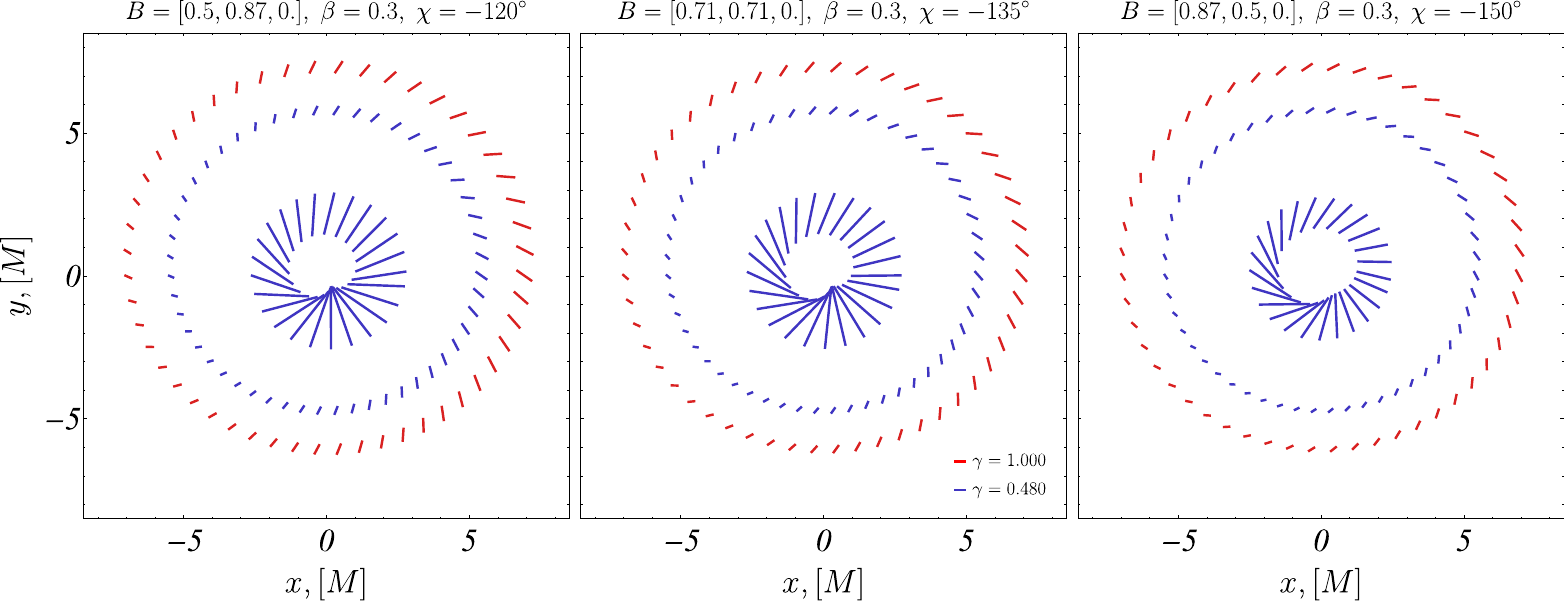} \caption{Polarized images for strongly naked singularities with scalar field parameter $\gamma=0.48$. We present in blue the observable polarization for the orbit located at $r=6M$ resulting in a couple of ring images. The polarization for the same orbit in the Schwarzschild spacetime is illustrated in red  for comparison. The inclination angle is $\theta = 20^\circ$.}
\label{fig:pol_sn}
\end{figure}

Here we examine the polarization structure of the double images of the equatorial circular orbits. For the purpose we simulate the observable polarization for photon trajectories emitted at radial coordinate $r=6M$ for a representative value of the scalar field parameter $\gamma = 0.48$. Our results are presented in Fig. $\ref{fig:pol_sn}$ for several magnetic field configurations compared to the polarized images of the same equatorial orbit for the Schwarzschild black hole. The polarization of outer image of the circular orbit at $R=6M$ for strongly naked singularities resembles the black hole case leading to a qualitatively similar pattern. In Table $\ref{table:theta20_sn}$ we evaluate the deviation in the polarization intensity between the two images by considering the ratio of the maximal values which is reached for the naked singularity and the black hole. We see that they deviate at most by $70\%$  for the magnetic field direction $B=[0.5,0.87, 0]$. 

The second image of the circular orbit for strongly naked singularities, which is located at small celestial angles, possesses rather different properties. It is characterized by a distinct polarization pattern around the ring with a fast variation of the EVPA in the lower part of the image. In addition, the polarization intensity grow up more than three times compared the  polarization for the Schwarzschild black hole (see Table $\ref{table:theta20_sn}$). Due to these properties the polarized flux from the inner image of the circular orbits provides significant observational signatures for distinguishing black holes from strongly naked singularities in the imaging experiments.


\begin{table}
    \centering
      \begin{tabular}{||c|c|c||}
       \hline
         \thead{ Magnetic field }   &  \thead{$\frac{ \text{max}\,\text{I}_\text{NS}}{\text{max}\,\text{I}_\text{Sch}}$\, (Outer ring)}& \thead{$\frac{ \text{max}\,\text{I}_\text{NS}}{\text{max}\,\text{I}_\text{Sch}}$\, (Inner ring)}
          \\  \hline

           \thead{\vspace{0.1mm}$\text{B = [0.5, 0.87, 0]}$\vspace{0.1mm}}  & \thead{0.7} &\thead{3.58} 
          \\  \hline

          \thead{\vspace{0.1mm}$\text{B = [0.71, 0.71, 0]}$\vspace{0.1mm}} &  \thead{0.6} & \thead{3.58}   
          \\  \hline

          \thead{\vspace{0.1mm}$\text{B = [0.87, 0.5, 0]}$\vspace{0.1mm}}  &  \thead{0.51} & \thead{2.89}  
          \\  \hline
       \end{tabular}
      \caption{Deviation of the polarization intensity for the strongly naked singularities from the Schwarzschild black hole. We evaluate the ratio of the maximal values of the polarization intensity which is reached for the naked singularity max $\text{I}_\text{NS}$ and  the Schwarzschild black hole max $\text{I}_\text{Sch}$ considering the outer and inner ring images. The scalar field parameter is $\gamma = 0.48$ and the inclination angle is $\theta=20^{\circ}$. }
    \label{table:theta20_sn}
\end{table}

\section{Conclusion}

In this work we study the linear polarization of the radiation from the accretion disk around the Janis-Newman-Winicour naked singularity. We apply an analytical model of a magnetized fluid ring orbiting in the equatorial plane and emitting synchrotron radiation and compute the observable polarized images in physical settings compatible with the radio source M87. The results are compared to the Schwarzschild black hole aiming to access the possibility to distinguish black holes from naked singularities by polarimetric experiments.

We investigate two classes of naked singularities with different lensing properties divided into weakly and strongly naked singularities. The direct polarized images of weakly naked singularities closely resemble the Schwarzschild black hole for small inclination angles. For equatorial magnetic fields the deviation in the polarization intensity between the two types of spacetimes is less than $7.5\%$, while the deviation in the polarization direction is less than $1.5\%$. Our previous studies on traversable wormholes showed similar results proving that it is difficult to put constrains on the nature of the compact object at the center of M87 by analyzing the direct polarized images.

More significant effects are observed if we consider larger inclination angles or higher order lensed images. For the inclination angle $\theta=70^\circ$ the polarization pattern still resembles the Schwarzshchild black hole, however the deviation in the polarization intensity and direction may reach $26.5\%$ and $6.8\%$ respectively for equatorial magnetic fields. The indirectly lensed images of order $k=1$ show modifications in the polarization pattern as the maximal intensity of the polarized flux concentrates at the upper part of the image. The polarization intensity may become four time larger compared to the Schwarzschild black hole for some magnetic field configurations, while the EVPA may increase up to three times for some azimuthal angles.

Strongly naked singularities lead to significant observational signatures already in the direct polarized images. They produce a second image of the fluid ring at small celestial angles with a characteristic twist of the observable polarization direction. The intensity of the polarized flux at the second image may become more than three times larger compared to the Schwarzschild black hole. These properties imply that the additional ring structure is observationally significant for polarimetric experiments and can be used for distinguishing black holes from strongly naked singularities by means of their polarization.

\section{Appendix}

\subsection*{A. Direct polarized images: examples}

\begin{figure}[t]
    \centering
   \includegraphics[width=\textwidth]{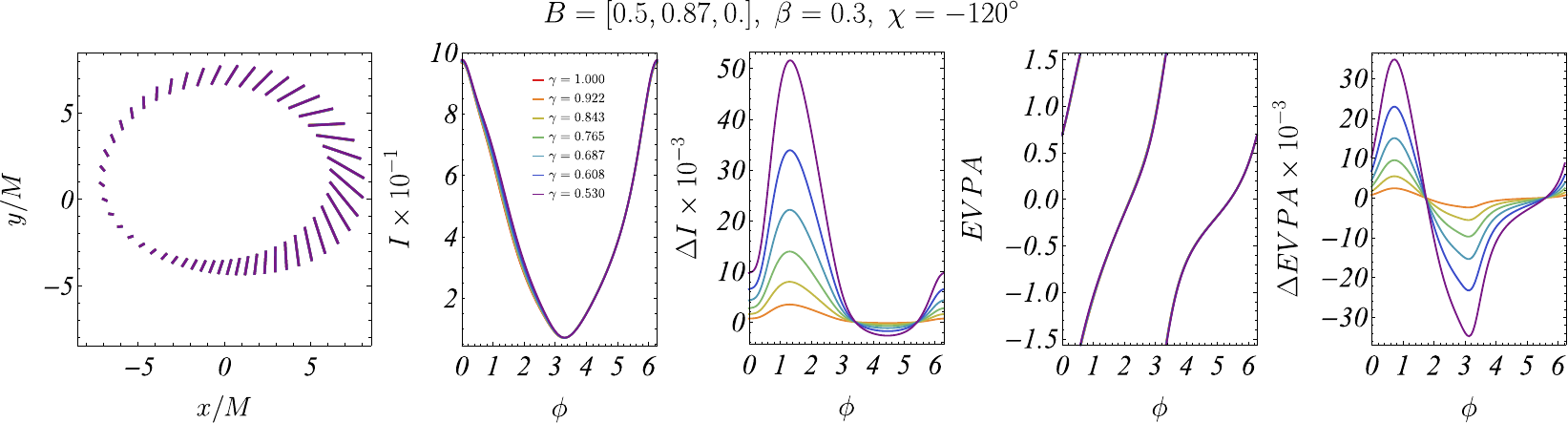}
 \\[2mm]
  \includegraphics[width=\textwidth]{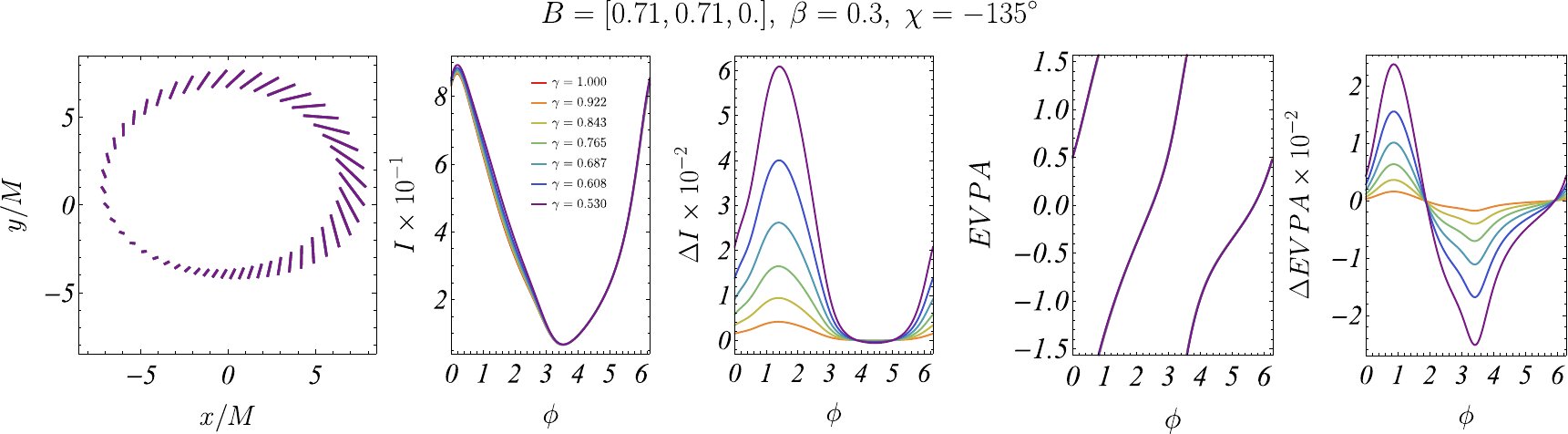} \\[2mm]
  \includegraphics[width=\textwidth]{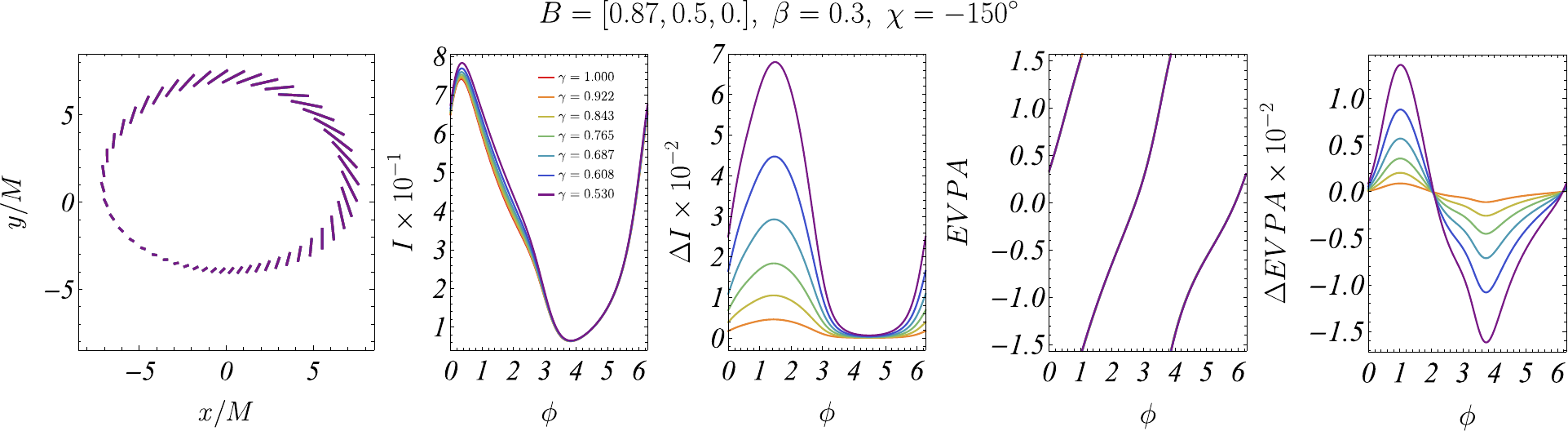}
 \caption{Linear polarization for weakly naked singularities at the inclination angle $\theta =50^\circ$. We analyze the polarization intensity I and direction EVPA as a function of the scalar field parameter $\gamma$, as well as their deviation from the Schwarzschild black hole $\Delta$I and  $\Delta$EVPA. }
\label{fig:polarization_50}
\end{figure}

We examine further the influence of the gravitational lensing on the observable polarization by considering additional inclination angles for weakly naked singularities. In Fig. $\ref{fig:polarization_50}$ we present the polarized images at the apparent location of the ISCO for the Schwarzchild black hole for the inclination agle of the observer $\theta = 50^\circ$. The polarization intensity and direction are compared to the Schwarzschild black hole for several values of the scalar field parameter and magnetic field configurations similar to the analysis performed in section 4.1. Table $\ref{table:theta50}$ summerizes the maximal deviations  of the polarization properties from black holes which are reached for our class of naked singularities.

\subsection*{B. Indirect polarized images: examples}

\begin{figure}
   \centering
   \includegraphics[width=\textwidth]{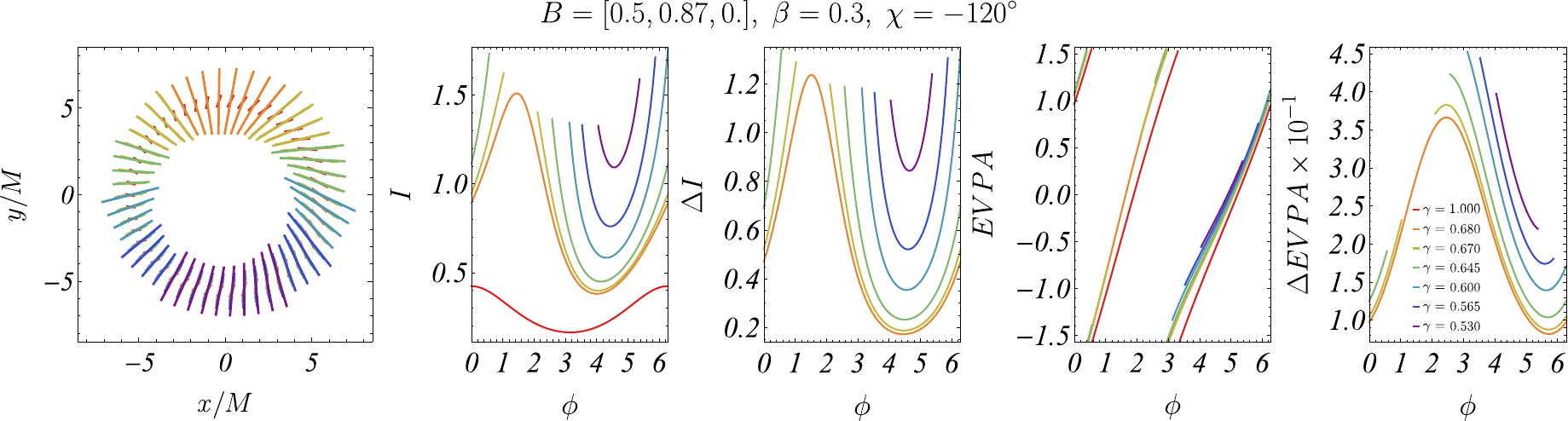}
 \\[2mm]
  \includegraphics[width=\textwidth]{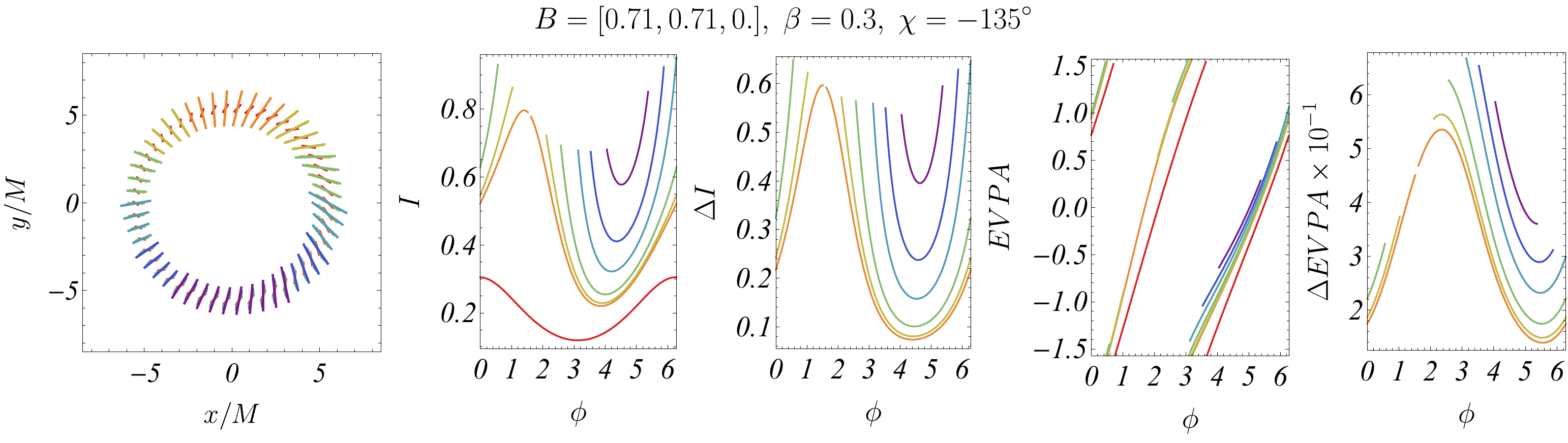} \\[2mm]
  \includegraphics[width=\textwidth]{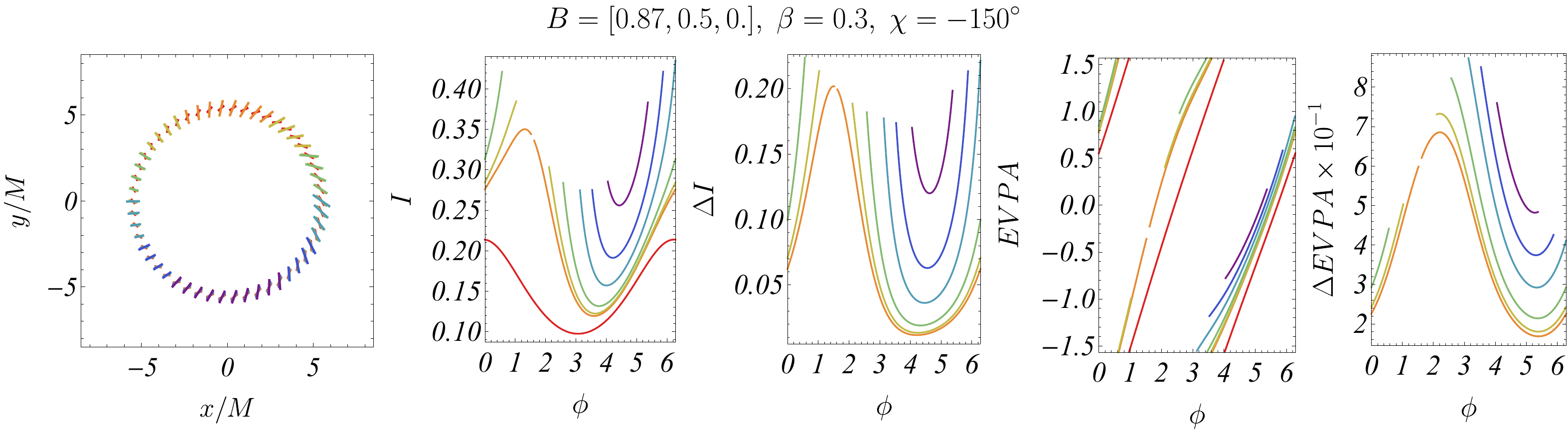}
  \caption{ Polarization of the indirect images of order $k=1$ for naked singularities with scalar field parameter in the range $\gamma\in$ at the inclination angle $\theta = 20^\circ$. We analyze the polarization intensity I and direction EVPA as a function of the redshift parameter $\alpha$, as well as their deviation from the Schwarzschild black hole $\Delta$I and $\Delta$EVPA.}
\label{fig:pol2_k1}
\end{figure}

We present the indirect images for the scalar fields $\gamma\in [0.53,0.68]$ which fail to produce a complete ring at the apparent location of the ISCO for the Schwarzschild black hole (see Fig. $\ref{fig:pol2_k1}$). For certain azimuthal angles there exist no geodesics originating at the equatorial plane in the naked singularity spacetime which get projected at the corresponding location in the observer's sky. In Table $\ref{table:theta20_2}$ we evaluate the maximum deviation in the polarization properties from the Schwarzschild black hole for the azimuthal angles where both compact objects produce images.

\section*{Acknowledgments}
We gratefully acknowledge support by the Bulgarian NSF Grant KP-06-H38/2.

\begin{table} 
    \centering
      \begin{tabular}{||c|c|c|c||}
       \hline
         \thead{ Magnetic field }   & \thead{$\left(\frac{\text{max}\,\Delta \text{I}}{\text{I}_\text{Sch}} \, [\%], \, \phi \, [rad]\right)$} & \thead{$\left(\frac{\text{max}\,\Delta \text{EVPA}}{\text{EVPA}_\text{Sch}} \, [\%] , \, \phi \, [rad]\right)$} & $\gamma$
          \\  \hline

           \thead{$\vspace{0.1mm}\text{B = [0.5, 0.87, 0]}$\vspace{0.1mm}}  &  \thead{(10.14, 0.425$\pi$)} & \thead{(3.064, 0.998$\pi$)} & \thead{0.530}
          \\  \hline

          \thead{\vspace{0.1mm}$\text{B = [0.71, 0.71, 0]}$\vspace{0.1mm}} &  \thead{(12.47, 0.453$\pi$)} & \thead{(1.972, 1.090$\pi$)} & \thead{0.530}
          \\  \hline

          \thead{\vspace{0.1mm}$\text{B = [0.87, 0.5, 0]}$\vspace{0.1mm}}  &  \thead{(14.27, 0.477$\pi$)} & \thead{(1.180, 1.190$\pi$)} & \thead{0.530}
          \\  \hline
          
       \end{tabular}
      \caption{Deviation of the polarization intensity and direction for the weakly naked singularities from the Schwarzschild black hole. In each case we give the maximum relative deviations $\text{max}\,\Delta\text{I}/\text{I}_\text{Sch}$ and $\text{max}\,\Delta\text{EVPA}/\text{EVPA}_\text{Sch}$  with respect to the Schwarzschild solution, which are reached in the polarized images, and the corresponding azimuthal angle.  The inclination angle is $\theta=50^{\circ}$.}
    \label{table:theta50}
\end{table}

\begin{table}
    \centering
      \begin{tabular}{||c|c|c||}
       \hline
         \thead{ Magnetic field }   & \thead{$\left(\frac{\text{max}\,\Delta \text{I}}{\text{I}_\text{Sch}} , \, \phi \, [rad]\right)$} & \thead{$\left(\frac{\text{max}\,\Delta \text{EVPA}}{\text{EVPA}_\text{Sch}}  , \, \phi \, [rad]\right)$} 
          \\  \hline

           \thead{\vspace{0.1mm}$\text{B = [0.5, 0.87, 0]}$\vspace{0.1mm}}  &  \thead{$(9.84, 2\pi)$} & \thead{$(0.59, 0.786\pi)$} 
          \\  \hline

          \thead{\vspace{0.1mm}$\text{B = [0.71, 0.71, 0]}$\vspace{0.1mm}} &  \thead{$(6.98, 2\pi)$} & \thead{$(2.25, 0.750\pi)$} 
          \\  \hline

          \thead{\vspace{0.1mm}$\text{B = [0.87, 0.5, 0]}$\vspace{0.1mm}}  &  \thead{$(3.46, 2\pi)$} & \thead{$(5.15, 0.822\pi)$} 
          \\  \hline

       \end{tabular}
      \caption{Deviation of the polarization for naked singularities with scalar field parameter in the range $\gamma\in [0.53, 0.68]$ from the Schwarzschild black hole for the indirect images of order $k=1$ and inclination angle $\theta=20^{\circ}$. In each case we give the maximum relative deviations $\text{max}\,\Delta\text{I}/\text{I}_\text{Sch}$ and $\text{max}\,\Delta\text{EVPA}/\text{EVPA}_\text{Sch}$  with respect to the Schwarzschild solution, which are reached in the polarized image.}
    \label{table:theta20_2}
\end{table}

\end{document}